\documentclass[12pt,letterpaper]{article}
\usepackage[utf8]{inputenc}
\usepackage{xcolor}
\usepackage{natbib}
\usepackage[colorlinks=true,linkcolor=black,anchorcolor=black,citecolor=black,filecolor=black,menucolor=black,runcolor=black,urlcolor=black]{hyperref}

\usepackage{comment}
\usepackage[framemethod=tikz]{mdframed}
\usetikzlibrary{shadows}
\definecolor{captiongray}{HTML}{555555}
\mdfsetup{%
  innertopmargin=2ex,
  innerbottommargin=1.8ex,
  linecolor=captiongray,
  linewidth=0.5pt,
  roundcorner=1pt,
  shadow=false,
}

\newlength{\figurewidth}
\sloppypar

\makeatletter
\renewcommand\section{\@startsection {section}{1}{\z@}%
  {-3.25ex \@plus -1ex \@minus -.2ex}%
  {1.5ex \@plus .2ex}%
  {\raggedright\normalfont\large\bfseries}}
\makeatother

\usepackage{amsmath, amssymb, mathtools, mathrsfs}
\newcommand{\floor}[1]{\left\lfloor #1 \right\rfloor}
\DeclareMathOperator{\FT}{\mathscr{F}}

\newcommand{\documentname}{\textsl{Note}}
\newcommand{\sectionname}{Section}
\newcommand{\foreign}[1]{\textsl{#1}}
\newcommand{\code}[1]{\texttt{#1}}

\begin{document}\thispagestyle{plain}

\section*{\raggedright Fitting very flexible models: Linear regression with large numbers of parameters%
\footnote{%
It is a pleasure to thank
Jed Brown (CU Boulder),
Dan Foreman-Mackey (Flatiron),
Alessandro Gentilini,
Teresa Huang (JHU),
Sam Roweis (deceased),
Adrian Price-Whelan (Flatiron),
Bernhard Sch\"olkopf (MPI-IS),
Kate Storey-Fisher (NYU),
Rachel Ward (UT Austin), and
Lily Zhao (Yale)
for valuable conversations and input.
SV is partially funded by NSF DMS 2044349, EOARD FA9550-18-1-7007, and the NSF--Simons Research Collaboration on the Mathematical and Scientific Foundations of Deep Learning (MoDL) (NSF DMS-2031985).}}

\noindent
\textbf{David W. Hogg} \\
\textsl{\footnotesize Flatiron Institute, a division of the Simons Foundation \\
Center for Cosmology and Particle Physics, Department of Physics, New York University \\
Center for Data Science, New York University \\
Max-Planck-Institut f\"ur Astronomie, Heidelberg}

\medskip
\noindent
\textbf{Soledad Villar} \\
\textsl{\footnotesize Department of Applied Mathematics \& Statistics, Johns Hopkins University \\
Mathematical Institute for Data Science, Johns Hopkins University}

\paragraph{Abstract:} There are many uses for linear fitting; the context here is interpolation and denoising of data, as when you have calibration data and you want to fit a smooth, flexible function to those data.
Or you want to fit a flexible function to de-trend a time series or normalize a spectrum.
In these contexts, investigators often choose a polynomial basis, or a Fourier basis, or wavelets, or something equally general.
They also choose an order, or number of basis functions to fit, and (often) some kind of regularization.
We discuss how this basis-function fitting is done, with ordinary least squares and extensions thereof.
We emphasize that it is often valuable to choose \emph{far more parameters than data points}, despite folk rules to the contrary:
Suitably regularized models with enormous numbers of parameters generalize well and make good predictions for held-out data; over-fitting is not (mainly) a problem of having too many parameters.
It is even possible to take the limit of infinite parameters, at which, if the basis and regularization are chosen correctly, the least-squares fit becomes the mean of a Gaussian process.
We recommend cross-validation as a good empirical method for model selection (for example, setting the number of parameters and the form of the regularization),
and jackknife resampling as a good empirical method for estimating the uncertainties of the predictions made by the model.
We also give advice for building stable computational implementations.

\section{Introduction}

In contexts in which we want to fit a flexible function to data, for interpolation or denoising, we often perform linear fitting in a generic basis, such as polynomials, Fourier modes, wavelets, or spherical harmonics.
This kind of linear fitting arises in astronomy when, for example, we want to calibrate the relationship between wavelength and position on the detector in a spectrograph:
We have noisy measurements of calibration data and we want to fit a smooth function of position that denoises and interpolates the calibration data.
It also arises when we want to make a data-driven interpolation, extrapolation, or local averaging of data, as with light-curve de-trending, continuum estimation, and interpolation or extrapolation of instrument housekeeping (or other) data.

When faced with problems of this kind, investigators have three general kinds of choices that they have to make:
They have to choose the basis in which they are working (Fourier, polynomial, wavelet, etc.).
They have to choose to what order they extend the basis---that is, how many components to use in the fit.
And they have to decide how (or whether) to regularize the fit, or discourage fit coefficients from getting out of line when the data are noisy or the basis functions are close to (or strictly) degenerate.

If you have encountered and solved problems like these, you have made these three kinds of choices (sometimes implicitly).
The second choice---about the number of coefficients to fit---is usually made heuristically, and often subject to the strongly believed opinion that you \emph{must have fewer parameters than data points}.
Here we are going to show that this folk rule is not valid; you can go to extremely large numbers of coefficients without trouble.
But like most folk rules, it has a strong basis in reality: There are extremely bad choices possible for the number of coefficients, and especially when the number of parameters is close or comparable to the number of data points.
As we will discuss below, these choices ought to be made with care.

In many cases the third kind of choice---about regularization---is made implicitly, not explicitly.
Here we are going to emphasize this choice and its importance, and its value in improving your results.

Alternatively, if you are unhappy with the three choices of basis, order, and regularization, you might decide to avoid such decisions and go fully \emph{non-parametric}:
Instead of fitting a basis expansion, you can use a strict interpolator (like a cubic spline), or you can fit a Gaussian process to your data.
Here we will show that the choice of any Gaussian process kernel function is equivalent to choosing a basis and a regularization and letting the number of fit components go to infinity.
That is, going non-parametric doesn't really get you out of making these choices.
It just makes these choices more implicit.
And it is a pleasure to note that any time you have gone non-parametric you have implicitly chosen to use a basis with way more fit parameters than data points!
The fact that non-parametrics work so well is strong evidence against the folk rule about the number of parameters needing to be less than the number of data points.

An important assumption or setting for the problems we are addressing in this \documentname\ will be that you care about predicting new data or interpolating the data, but you explicitly \emph{don't care} about the parameters of the fit or the weights of the basis functions \foreign{per se}.
In this setting there are no important meanings to the components of the model.
That is---for us---only the data exist.
The details of the model are just choices that permit high-quality interpolations and predictions in the space of the data.

In what follows, we will look at applications that look like interpolation; there are many other contexts in which regressions have gone to very large numbers of parameters.
For example, there are contexts in which there are enormous numbers of possible natural features, like for instance when the data are images or videos: Every pixel of every frame of the video---or any linear (or nonlinear) combination of pixels---can become a feature for the regression.
Also, there are contexts in which features are generated randomly from, say, a space of functions that can act on the natural features (\citealt{rahimi2007random}).
These settings are not what we are addressing here, but they are relevant and related
(many good books exist, for example, \citealt{bishop}, \citealt{esl}, \citealt{agresti}, \citealt{gelman}).
In some ways, the most flexible of models currently in use are deep networks, where it is both the case that the input data often have enormous numbers of natural features, \emph{and} the deep network is capable of generating (effectively) far more, internally.

In some contexts you have strong beliefs about the noise affecting your measurements. In other cases you don't.
In some cases you have strong reasons to use a particular basis.
In other cases you don't.
The differences in these beliefs and the differences in your objectives will change what methods you choose and how you use and analyze them.
We'll try to be useful to you no matter where you're at.
This document does not deliver new research results on the mathematics or statistics of regression.
It is novel only in that it makes very specific the connection between regularized linear regression and Gaussian processes.

Ordinary least squares is reviewed in \sectionname~\ref{sec:ols}.
The extensions of weighted least squares and ridge regression are shown in \sectionname~\ref{sec:extensions}.
The over-para\-meterized case (more parameters than data) is discussed in \sectionname~\ref{sec:overpar}.
The concept of feature weighting for controlling regularization in over-parameterized fits is introduced in \sectionname~\ref{sec:fwols}.
Cross-validation is explained and used to choose the number of parameters in \sectionname~\ref{sec:dd}, and the double descent phenomenon is shown.
The Gaussian Process appears as the limit of infinite parameters in \sectionname~\ref{sec:gp}.
Jackknife resampling is explained and used to estimate uncertainties in \sectionname~\ref{sec:uncertainty}.
Numerical implementation considerations are discussed in \sectionname~\ref{sec:implementation} and some final remarks are made in \sectionname~\ref{sec:discussion}.

\section{Standard linear fitting: Ordinary least squares with a feature embedding}\label{sec:ols}

Our setup will be that there are $n$ scalar data points $y_i$.
Each of these data points has an associated coordinate or location $t_i$.
In the machine-learning lexicon, these will be our ``training data''.
The location $t_i$ could be thought of as a time at which the data point was taken, or a position, or it can be a higher dimensional vector or blob of housekeeping data associated with the data point.
Critically, we are going to imagine that the $t_i$ are known very well (not very uncertain or noisy), while the $y_i$ are possibly very uncertain or noisy measurements.
We'll return to these assumptions at the end.

We are going to fit these data $y_i$ with a linear sum of $p$ basis functions $g_j(t)$. These basis functions are functions of the coordinates $t$. That is, our implicit generative model is
\begin{equation}
    y_i = \sum_{j=1}^p \beta_j\,g_j(t_i) + \mathrm{noise}
    ~,
\label{eq.model}
\end{equation}
where the $p$ values $\beta_j$ are parameters or coefficients of the linear fit. We can assemble the evaluations of the $p$ functions $g_j(t)$ at the $n$ data coordinates $t_i$ into a $n\times p$ design matrix or feature matrix $X$ such that
\begin{equation}
    [X]_{ij} = g_j(t_i)
    ~.
\end{equation}
The transformation of the $n$ locations $t_i$ into a $n\times p$ matrix $X$ is the opposite of a dimensionality reduction.
It is called variously a \emph{feature map} or a \emph{feature embedding}.
We like ``embedding'' because it (almost always) raises the dimensionality of the locations $t$ into the $p$-dimensional space of the rows of $X$.

\begin{figure}[t]
    \begin{mdframed}
    \includegraphics[width=\figurewidth]{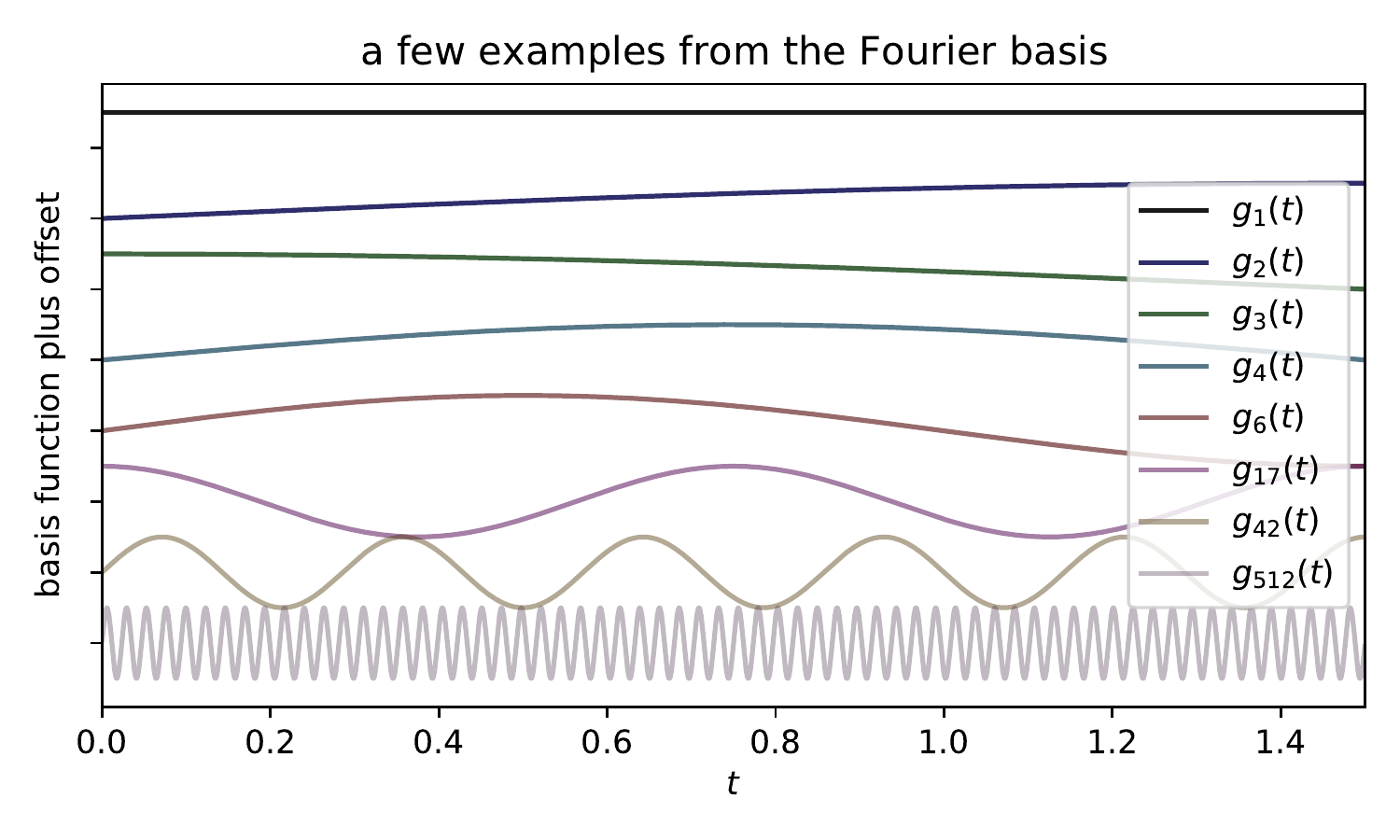}
    \caption{Examples of basis functions $g_j(t)$ from the basis given in equation~\eqref{eq:basis}. This basis was constructed with length-scale parameter $T=3$. The wavelength in the location space decreases, and the frequency increases, with index $j$.}
    \label{fig:basis}
    \end{mdframed}
\end{figure}

For a concrete example, one common choice is to make the feature embedding functions $g_j(t)$ terms in a Fourier series
\begin{align}\label{eq:basis}
    g_j(t) &= \left\{\begin{array}{ll}
            \cos\omega_j\,t & \mbox{for $j$ odd} \\
            \sin\omega_j\,t & \mbox{for $j$ even}\end{array}\right.
    \\
    \omega_j &= \frac{\pi}{T}\,\floor{\frac{j}{2}}
    ~,
\end{align}
where $T$ is a (large) length-scale in the coordinate space ($t$ space) and $\floor{j/2}$ indicates the floor of $j/2$ (integer division).
Example functions $g_j(t)$ from this basis are shown in \figurename~\ref{fig:basis}.
Alternative common choices would be to make the embedding functions $g_j(t)$ polynomials or other kinds of ordered basis functions, such as wavelets or spherical harmonics (the latter if, say, the $t_i$ are positions on the sphere).
Another choice that isn't common in the natural sciences, but studied in machine learning (for example, \citealt{rahimi2007random}), is to choose the features randomly from a distribution (rather than on a regular grid in frequency, as we do here). That is beyond our scope, but we come back to it in \sectionname~\ref{sec:discussion}.

The idea of least-squares fitting is that the ``best'' values of the parameters $\beta_j$ are the values that minimize the sum of squares of the differences between the data and the linear combination of features:
\begin{equation}\label{eq:opt1}
    \hat{\beta} = \arg\min_\beta \|Y - X\,\beta\|_2^2
    ~,
\end{equation}
where $\hat{\beta}$ is the $p$-vector (column vector) of the $p$ best-fit values $\hat{\beta}_j$ of the parameters $\beta_j$, and $\|q\|_2^2$ denotes the squared L2-norm, or the \emph{sum of squares} of the components of a vector $q$
\begin{equation}
    \|q\|_2^2 \equiv q^\top q
    ~.
\end{equation}
This optimization objective \eqref{eq:opt1} is convex and the optimization problem has a solution in closed form\footnote{
Note that $\nabla_{\beta} \|Y-X\beta\|^2 = 2X^\top(Y-X\beta)$ so critical points are such that $X^\top Y= X^\top X\beta$. Since the objective is convex all these critical points are global minima. Further analysis is given in Appendix \ref{app:math}.
} as long as the number of parameters (and features) $p$ is less than the number of training points $n$ (and the matrix $X^\top X$ is invertible):
\begin{equation}
    \hat{\beta} = (X^\top X)^{-1}\,X^\top Y
    ~.
    \label{OLS-under}
\end{equation}
We will treat the case in which the matrix is not invertible below.
When the investigator knows uncertainties on the training data $y_i$, this expression will change a bit; we begin to discuss that in the next \sectionname.

\begin{figure}[t]
    \begin{mdframed}
    \includegraphics[width=\figurewidth]{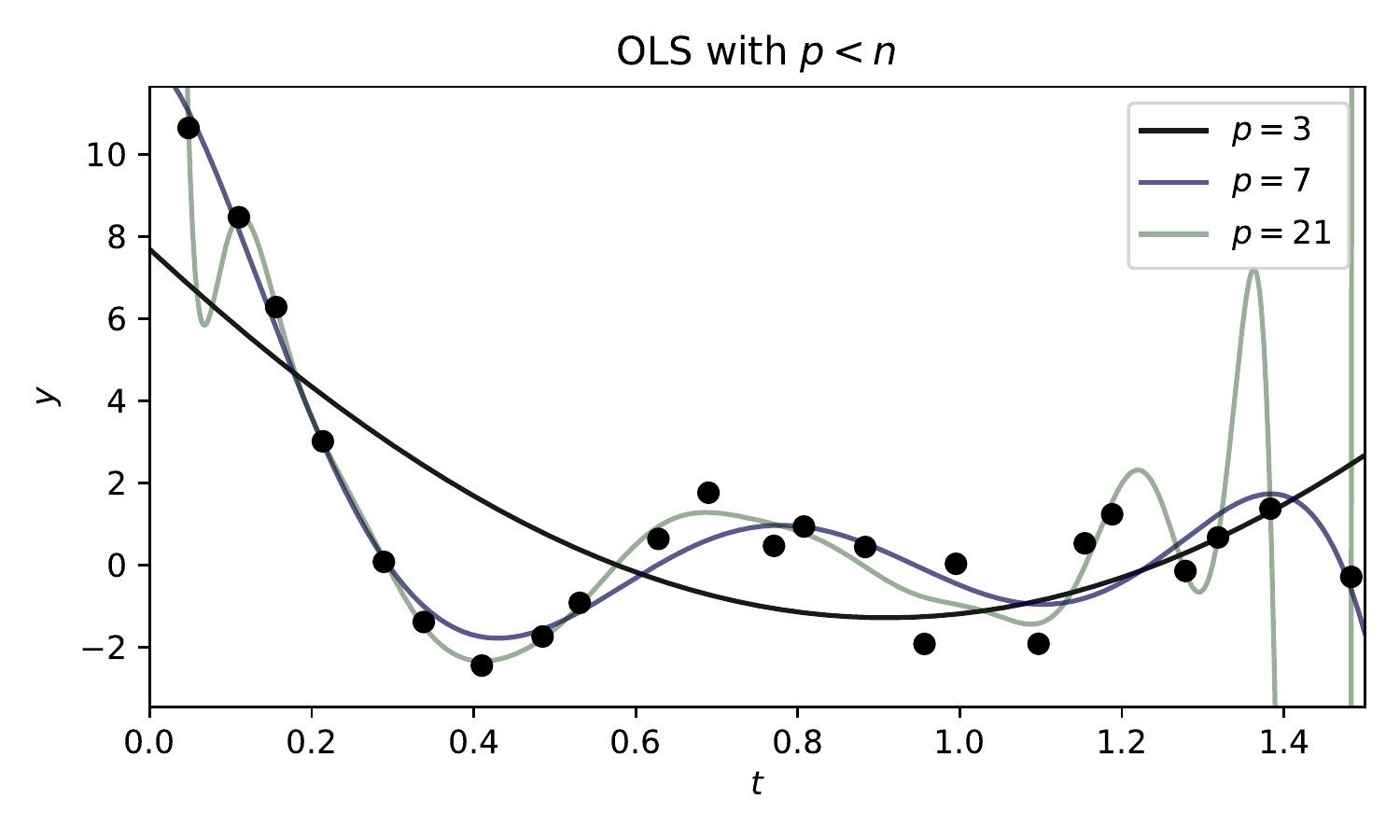}
    \caption{Ordinary least-squares (OLS) fits (continuous lines) to a set of example data points (black dots). Fits are shown for different values of the number of basis functions $p$; there are $n=23$ data points. Here (and in all the \figurename s to follow) we are using the Fourier basis functions in \eqref{eq:basis} and shown in \figurename~\ref{fig:basis}. The data $y_i$ were generated using a function that does not reside in the function space spanned by the basis. Here the $X_\ast$ matrices used in the predictions $\hat{Y}_\ast$ were generated from a fine grid of locations in the location coordinate $t$; the plots of the fine grid of predictions are the continuous lines. The fits with larger $p$ have more flexibility to fit the data than the fit with $p=3$, but the fit at the highest $p$ shows evidence of over-fitting.}
    \label{fig:ols1}
    \end{mdframed}
\end{figure}

But recall our setting:
We are using the linear fit to interpolate the data, or de-noise the data, or predict new data.
In these contexts, we don't care about the parameter vector $\hat{\beta}$ itself.
We care only about the predictions at a set of new ``test'' locations $t_\ast$, which will usually be different from the training locations $t_j$.
From the new test times $t_\ast$ we create the test feature matrix $X_\ast$ (by the same feature embedding functions $g_j(t)$).
The prediction $\hat{Y}_\ast$ for the $y$ values at the test locations $t_\ast$ becomes
\begin{equation}\label{eq:OLS}
    \hat{Y}_\ast = X_\ast\,(X^\top X)^{-1}\,X^\top Y
    ~.
\end{equation}
Examples of OLS applied to some toy data\footnote{The algorithm by which the toy data were made---and indeed all of the code used to make the figures for this \documentname---is available online at \url{https://github.com/davidwhogg/FlexibleLinearModels}.} are shown in \figurename~\ref{fig:ols1}.
This form \eqref{eq:OLS} of the prediction of new test data is called ordinary least squares (OLS). It has many good properties, some of which are encoded in the Gauss--Markov theorem.
In particular, if the noise contributions in the model given in equation \eqref{eq.model} are uncorrelated, have zero mean, and equal variances for $i=1,\ldots, n$ then the Gauss--Markov theorem states that the OLS estimator has the lowest variance within the class of unbiased estimators that are linear in $Y$ (see, for example, \citealt{esl}, Ch.~3).

\section{Discussion and extensions of OLS}\label{sec:extensions}

The prediction \eqref{eq:OLS} when $p<n$ (the under-parameterized or traditional regime) is \emph{affine invariant} in that $p$-dimensional rotations or rescalings of the rectangular feature matrix $X$ do not affect predictions.
That is, if $R$ is an invertible $p\times p$ matrix, the prediction using $X'\leftarrow X\,R$ will be identical to the prediction using the original $X$.
This affine invariance will be modified in the over-parameterized regime, below.

Although the OLS prediction \eqref{eq:OLS} is affine invariant with respect to $p$-dimensional transformations, it is \emph{not} affine invariant with respect to $n$-dimensional transformations, such as a re-weighting of the input data. Indeed, if you know weights or inverse variances for your data points, conceptually you can put them into a weight matrix $C^{-1}$ (written this way to emphasize that reweighting is usually inverse-variance weighting) and write
\begin{equation}\label{eq:wls}
    \hat{Y}_\ast = X_\ast\,(X^\top C^{-1}\,X)^{-1}\,X^\top C^{-1}\,Y
    ~,
\end{equation}
where the weight matrix $C^{-1}$ is $n\times n$ (and often diagonal in standard applications). The weight matrix $C^{-1}$ is sometimes called the information tensor (or information matrix) and its inverse $C$ is often called the covariance matrix or the noise variance tensor.

\begin{figure}[t]
    \begin{mdframed}
    \includegraphics[width=\figurewidth]{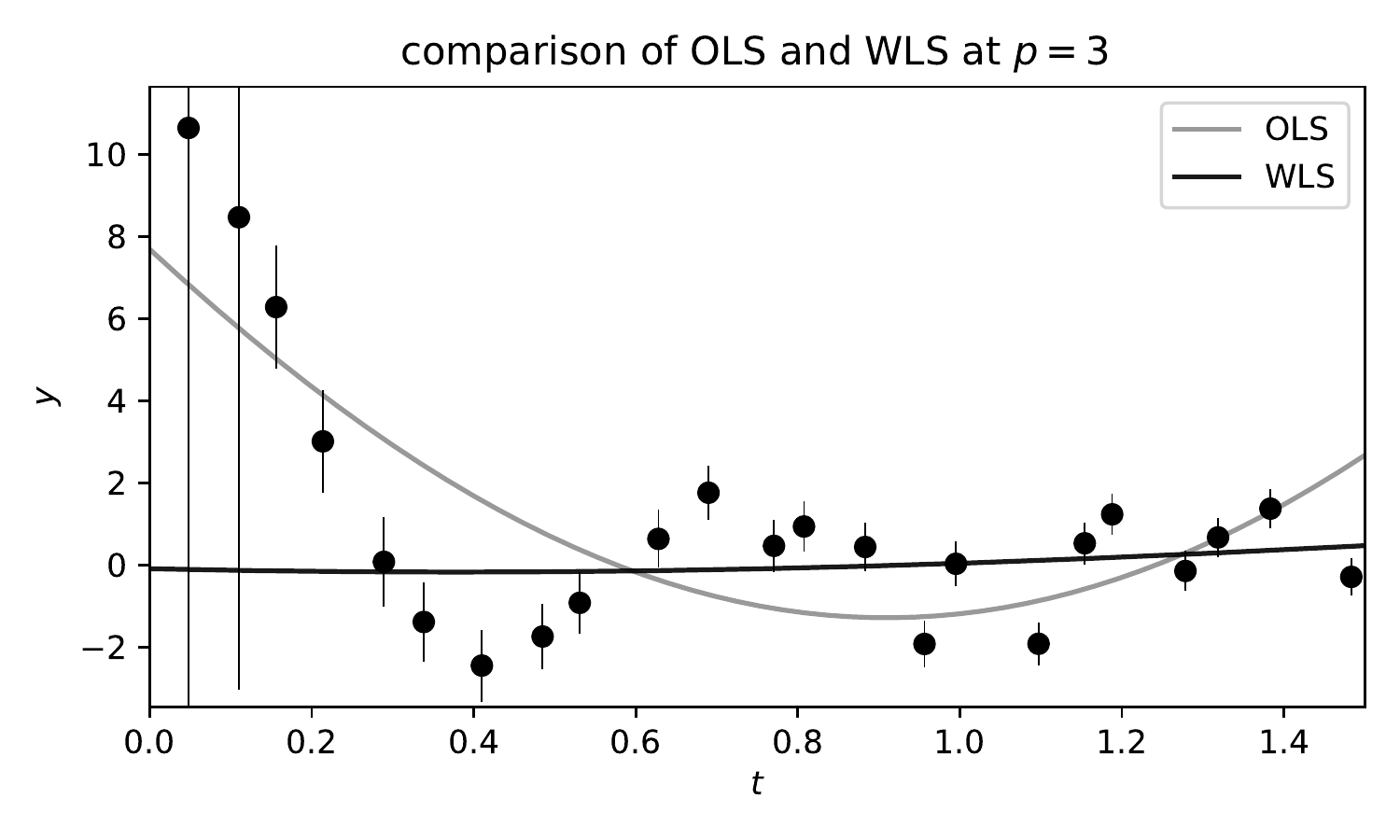}
    \caption{Comparison of ordinary least-squares (OLS) and weighted least-squares (WLS) fits (continuous lines) to the example data (black dots). In order to illustrate the differences, we assigned non-trivial error bars to the data. The error bars are ignored in the OLS fit, but in the WLS fit, the weight matrix $C^{-1}$ is diagonal with the diagonal entries set to the inverses of the squares of those error bars. The WLS fit ``pays less attention to'' the points on the left with the largest error bars.}
    \label{fig:wls}
    \end{mdframed}
\end{figure}

This form \eqref{eq:wls} of least squares is called \emph{weighted least squares} (WLS) because of the data weighting (not to be confused with \emph{feature} weighting, to appear below).
It is also called \emph{chi-squared fitting} because it optimizes the scalar objective commonly called chi-squared:
\begin{equation}
    \chi^2 = (Y - X\,\beta)^\top C^{-1}\,(Y - X\,\beta)
    ~.
\end{equation}
(If that isn't obviously chi-squared to you, recall that $C^{-1}$ is often diagonal and has the inverses of the squares of the data uncertainties on that diagonal.)
In \figurename~\ref{fig:wls} a comparison of OLS and WLS is shown, for a case of non-trivial data weights, where the data weights are set to be the inverse squares of individual data-point uncertainties.

We will say more about noisy data and the propagation of uncertainties below in \sectionname~\ref{sec:uncertainty}.
It might be crossing your mind that there are uncertainties not just in the data points $y_i$, but also often in the \emph{locations} $t_i$ of the data as well. It turns out that taking the latter into account is a much harder problem; we will discuss this briefly in \sectionname~\ref{sec:uncertainty}.

\begin{figure}[t]
    \begin{mdframed}
    \includegraphics[width=\figurewidth]{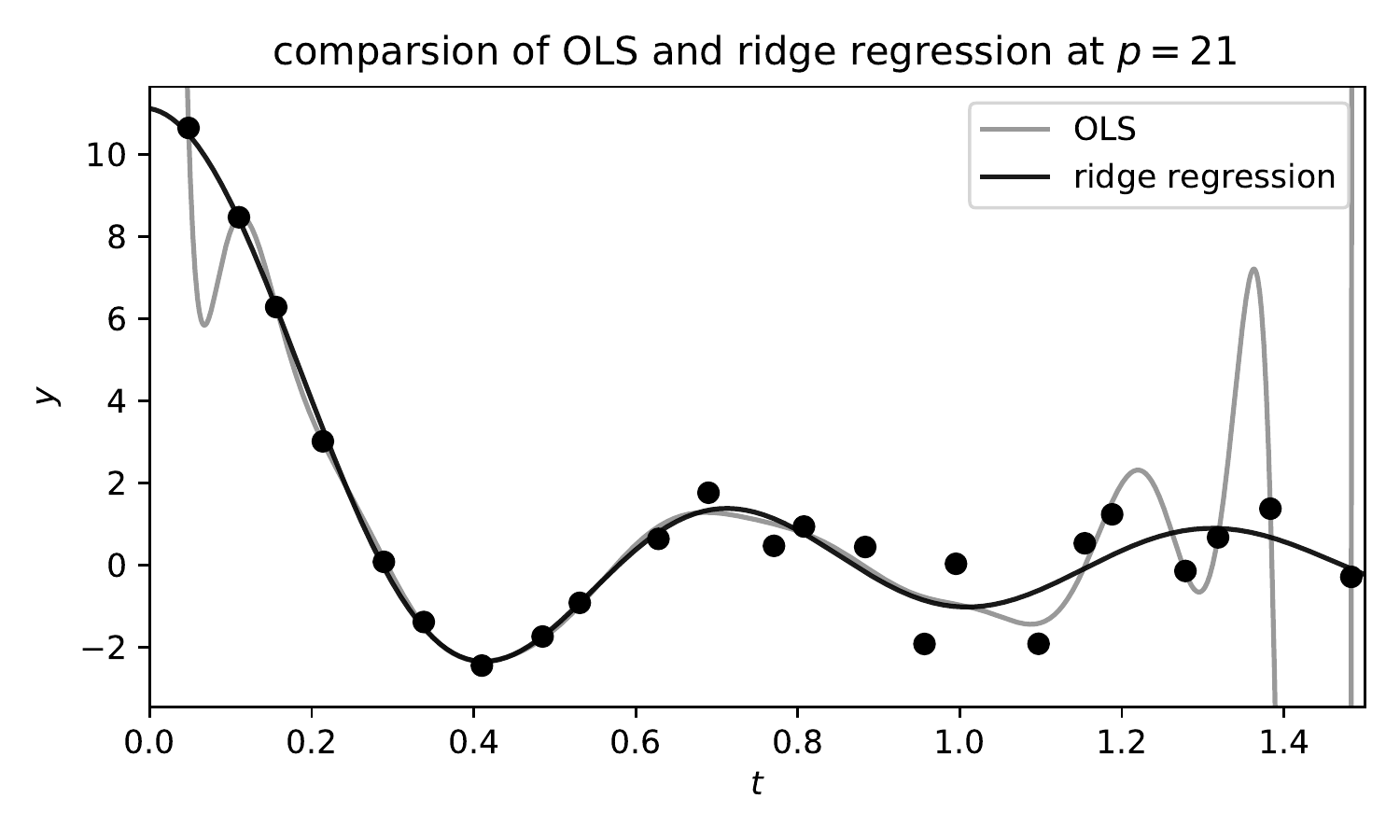}
    \caption{Comparison of ordinary least-squares (OLS) and ridge-regression fits (continuous lines) to the example data (black dots). In order to illustrate the differences, we chose the $p=21$-parameter fit, which shows evidence of over-fitting at the edges of the fit range. The regularized fit looks more sensible, though it fits the individual data points less precisely. The choice of regularization parameter $\lambda$ matters; here we used $\lambda=0.1$ (chosen heuristically by hand).}
    \label{fig:ridge}
    \end{mdframed}
\end{figure}

It is common to include a regularization that discourages the fit from making use of large amplitudes $\beta_j$.
There are many options, but the simplest is ridge regression (or Tikhonov regularization or L2 regularization), which (in the form that doesn't have data weights) looks like
\begin{equation} \label{eq:ridge}
    \hat{\beta} = \arg\min_\beta \|Y - X\,\beta\|_2^2 + \lambda\,\|\beta\|_2^2
    ~,
\end{equation}
where $\lambda>0$ is a regularization parameter that penalizes large values for elements $\beta_j$ of the parameter vector.
This optimization is also convex.
The ridge-regularized prediction for new data looks like
\begin{equation} \label{eq:ridge_sol}
    \hat{Y}_\ast = X_\ast\,(X^\top X + \lambda\,I)^{-1}\,X^\top Y
    ~,
\end{equation}
where $I$ is the $p\times p$ identity (see Appendix \ref{app:math}).
In the language of Bayesian inference, $\lambda$ can be seen as the inverse of a prior variance for the parameters $\beta_j$.
The salutary effect of the ridge regularization is shown in \figurename~\ref{fig:ridge}.

The ridge brings with it a choice: How to set the hyper-parameter $\lambda$?
We generally recommend cross-validation, to be discussed below in \sectionname~\ref{sec:dd}.
Ridge regression is not affine invariant in the sense that the standard OLS prediction \eqref{eq:OLS} is; that is, the effect of the regularization depends on the amplitudes and linear combinations of features placed in the feature matrix.
This will become important later when we consider feature weights below.
It is also the case that the regularization need not be proportional to the identity matrix: In principle any positive definite matrix $\Lambda$ could be used in place of $\lambda\,I$; this makes sense to consider when you have detailed prior beliefs about all the parameters $\beta_j$ or if those parameters are measured with different units (say).

You can combine both the point weighting from WLS and the generalized ridge regression into a weighted ridge that looks like
\begin{equation}\label{eq:foo}
    \hat{Y}_\ast = X_\ast\,(X^\top C^{-1}\,X + \Lambda)^{-1}\,X^\top C^{-1}\,Y
    ~.
\end{equation}
This form has good properties for many real-world physics applications, where data-point error bars are often known, and functions are often expected to be smooth.
We'll discuss this form \eqref{eq:foo} more below in \sectionname~\ref{sec:fwols},
but briefly we can say here that it appears in Bayesian inference contexts where the data points $y_i$ are treated as having Gaussian noise associated with them (with known variance tensor or covariance matrix $C$) and there is a Gaussian prior on the parameter vector $\beta$ (with known variance $\Lambda^{-1}$).
We have discussed that model elsewhere (\citealt{products}).\footnote{In our previous discussion of this product of Gaussians, the notation differs. What's called $\Lambda^{-1}$ here is called $\Lambda$ in \citet{products}.}
It is also useful sometimes to think of the ``units'' or dimensions of the quantities in \eqref{eq:foo}: In the Bayesian setting, the units of $C$ would be the square of the units of $Y$ and the units of $\Lambda^{-1}$ would be the square of the units of the ratio $Y/X$ (the square of the units of $\beta$).

\section{Over-parameterization}\label{sec:overpar}

We are taught folklore, at a young age, that we can never fit for more parameters than we have data. That is, we can never work at $p>n$.
This isn't true!
Not only is it the case that we \emph{can} work at $p>n$, in many cases we \emph{should} work at $p>n$, and many real-world regressions do.
But it is true that the $p>n$ regime is indeed strange!
In the over-parameterized case, there are typically many settings of the parameters $\beta_j$ that will literally zero out the differences between the data $Y$ and the linear prediction $X\,\beta$.
The OLS solution, in this case, is defined (somewhat arbitrarily) to be the minimum-norm parameter vector $\beta$ that interpolates the data:
\begin{equation}\label{eq:opt2}
    \hat{\beta} = \arg\min_\beta \|\beta\|_2^2 ~~\mbox{subject to}~~ Y = X\,\beta
    ~.
\end{equation}
Technically this formulation depends on an additional assumption that the feature matrix $X$ is full rank or that the data $Y$ lie in the subspace spanned by $X$.
This is true almost always when $p>n$.
This optimization is again convex and has a unique solution, although that solution will depend on feature weights that we discuss in a moment.
When the investigator knows uncertainties on the training data points $y_i$, this expression will change a bit; we return to that in the next \sectionname.
The under-parameterized and over-parameterized optimization statements \eqref{eq:opt1} and \eqref{eq:opt2} can be unified into one form by considering the limit of light L2 regularization:
\begin{equation}\label{eq:opt3}
    \hat{\beta} = \lim_{\lambda\to 0^+}\left[\arg\min_\beta \|Y - X\,\beta\|_2^2 + \lambda\,\|\beta\|_2^2\right]
    ~;
\end{equation}
in the limit, this delivers the OLS solution in either case and doesn't require the constraint in \eqref{eq:opt2} to be satisfied.

\begin{figure}[t]
    \begin{mdframed}
    \includegraphics[width=\figurewidth]{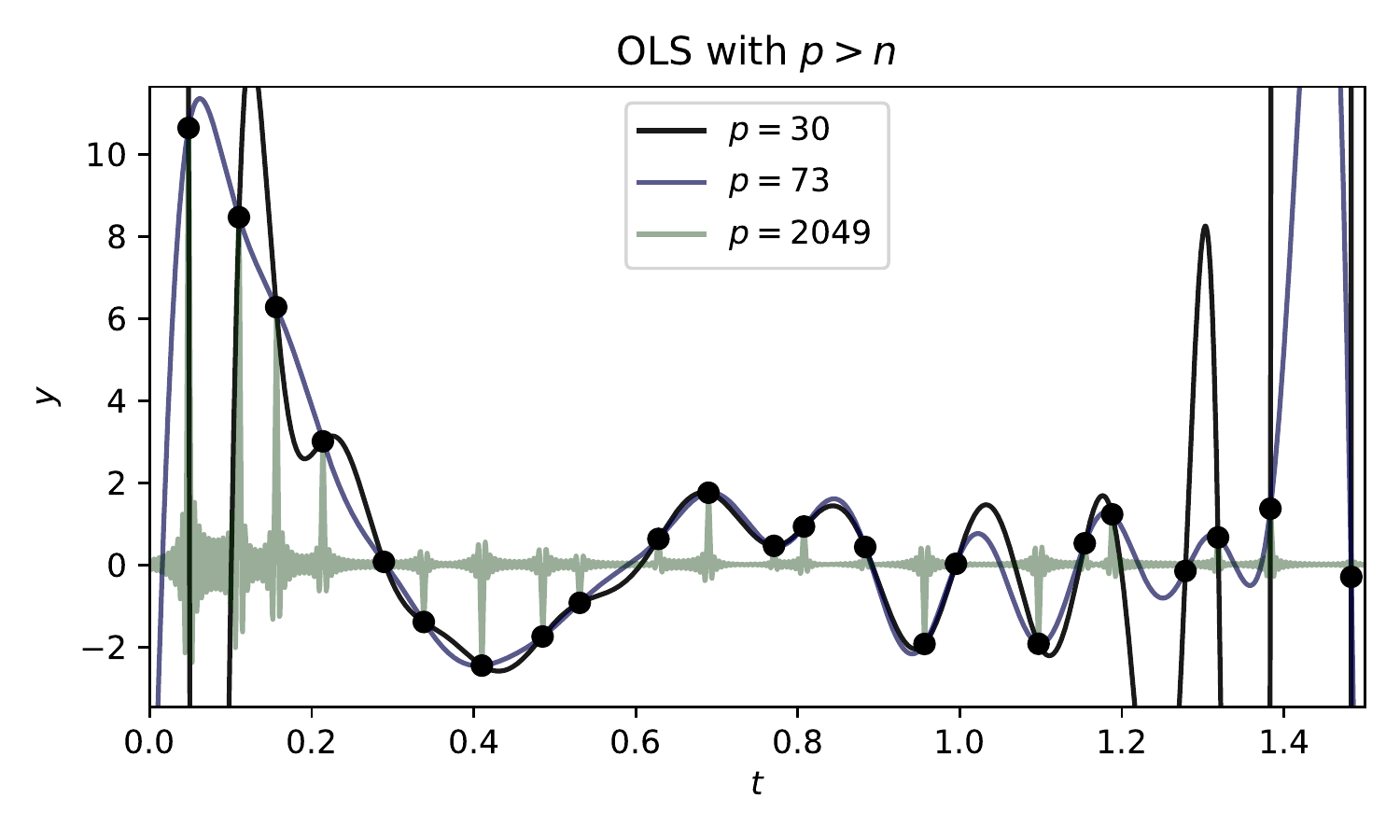}
    \caption{Ordinary least-squares (OLS) or min-norm least-squares fits (continuous lines) to a set of example data points (black dots), but now for a few over-parameterized cases. There are $n=23$ data points. As in \figurename~\ref{fig:ols1}, the data points $y_i$ were generated using a function that does not precisely reside in the function space spanned by the basis. The three fits are very different, but they all go through all the data points exactly. As $p$ gets large, the fit function approaches $y=0$ almost everywhere; this is a consequence of Plancharel's Theorem (see text).}
    \label{fig:ols2}
    \end{mdframed}
\end{figure}

In the over-parameterized case ($p>n$), the prediction looks like
\begin{equation}\label{eq:OLS2}
    \hat{Y}_\ast = X_\ast\,X^\top (X\,X^\top)^{-1}\,Y
    ~,
\end{equation}
provided that $X\,X^\top$ is invertible (which will usually be the case).
Like \eqref{eq:OLS}, this prediction is \emph{also} called ordinary least squares (OLS), or sometimes ``min-norm least-squares'' to emphasize the point that it is making the minimum-norm choice of $\beta$ among many degenerate solutions.
Examples of OLS fits in the over-parameterized regime are shown in \figurename~\ref{fig:ols2}.
Note that the fits with different numbers of parameters $p$ lead to very different predictions, but they all go through the data exactly.

It is slightly off-topic to notice that in \figurename~\ref{fig:ols2}, as $p$ gets very large, the OLS solution approaches $y=0$ everywhere that it can.
This behavior is a direct consequence of Plancharel's theorem, which states that the Fourier transform is unitary (see for instance \citealt{folland2016course}).
This means that the integral over frequency of the square of the Fourier transform is equal to the integral over location of the square of the original function.
Thus the min-norm solution delivered by OLS, which chooses the interpolating function that minimizes the squares of the component amplitudes $\beta$ in the Fourier basis used to make \figurename~\ref{fig:ols2}, will choose the interpolating function $y(t)$ that minimizes the mean square of the value of the function in the location space.
It will try to stay as close to $y=0$ as possible.
That's probably not desirable in most applications! We will fix that problem in the next \sectionname.

The two equations for OLS---\eqref{eq:OLS} and \eqref{eq:OLS2}---can be unified into one equation (and also generalized to handle non-invertible matrices $X^\top X$ and $X\,X^\top$) if we define the pseudo-inverse $X^\dagger$:
\begin{equation}\label{eq:OLS3}
    \hat{Y}_\ast = X_\ast\,X^\dagger\,Y
    ~.
\end{equation}
The pseudo-inverse of a diagonal matrix is defined as the diagonal matrix made by inverting the non-zero diagonal entries. And for any non-diagonal (or any non-square) matrix $X$ the pseudo-inverse is defined by taking the singular-value decomposition (SVD) of $X$, $X=U\,S\,V$, with $S$ diagonal and $U,V$ orthogonal, then $X^\dagger \equiv V^\top S^\dagger\,U^\top$.

This pseudo-inverse form \eqref{eq:OLS3} of OLS is extremely general: It works for both the $p\le n$ and $p>n$ cases, and it works when the $X\,X^\top$ or $X^\top X$ matrices are not invertible.
There can be significant numerical issues with implementing the pseudo-inverse; we comment on those in \sectionname~\ref{sec:implementation}.

\section{Feature weighting}\label{sec:fwols}

The OLS prediction for $p>n$ is \emph{not} affine invariant with respect to $p$-dimensional rotations or rescalings.
That is, rotations and scalings in the feature space \emph{will} affect predictions.
It behooves us to re-scale the features (the $p$ $n$-vectors of the feature matrix $X$) in a sensible way, like for instance, to encourage the fit to use low-frequency features more than high-frequency features (\citealt{xie2020weighted}, \citealt{bah2016sample}, \citealt{rauhut2016interpolation}).
We can encode these feature weights in a $p\times p$ diagonal weight matrix $\Lambda^{-1}$ and the prediction becomes
\begin{equation} \label{eq:weighted_sol}
    \hat{Y}_\ast = X_\ast\,\Lambda^{-1}\,X^\top (X\,\Lambda^{-1}\,X^\top)^{-1}\,Y
    ~.
\end{equation}
This can be seen as the prediction for new data resulting from the following optimization (again, assuming the data can be interpolated):
\begin{equation} \label{eq:weighted}
    \hat{\beta} = \arg\min_\beta \|\Lambda^{1/2}\,\beta\|_2^2 ~~\mbox{subject to}~~ Y = X\,\beta
    ~,
\end{equation}
which---in analogy to the optimizations \eqref{eq:opt2} and \eqref{eq:opt3}---can also be written as
\begin{equation}
    \hat{\beta} = \lim_{\lambda\to 0^+}\left[\arg\min_\beta \|Y - X\,\beta\|_2^2 + \lambda\,\|\Lambda^{1/2}\,\beta\|_2^2\right]
    ~.
\end{equation}
This doesn't require the constraints in \eqref{eq:weighted} to be satisfied. 
This optimization\footnote{If you are a physicist and you don't like the mathematical $\|\cdot\|_2^2$ notation---if you prefer to look at quantities that are obvious scalar forms---then recall that $\|\Lambda^{1/2}\,\beta\|_2^2 = \beta^\top\Lambda\,\beta$.
That right-hand-side object is a gauge-invariant object so long as $\beta$ and $\Lambda$ are also gauge-invariant objects themselves.
Yes, data analysis can have this kind of geometric structure!}
penalizes more strongly the parameters $\beta_j$ corresponding to features $g_j(t)$ with larger values of $[\Lambda^{1/2}]_{jj}$.

\begin{figure}[t]
    \begin{mdframed}
    \includegraphics[width=\figurewidth]{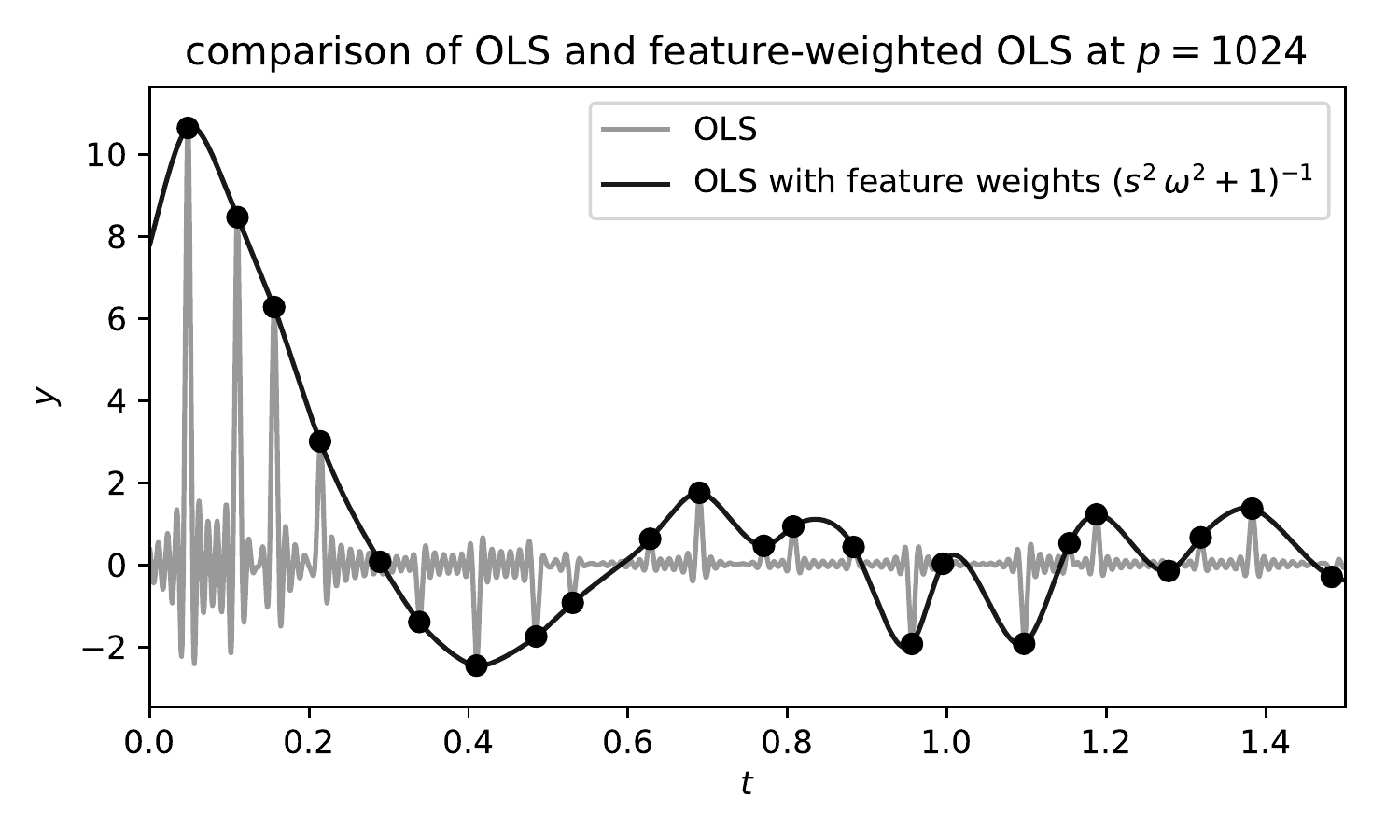}
    \caption{Comparison of ordinary least squares (OLS) to OLS with a specific feature weighting. The feature-weighting function is given in the text in equations~\eqref{eq:f} and \eqref{eq:lambdainv}; in this and the following \figurename s, we (somewhat arbitrarily) set the width parameter to be $s=0.05$. Because this weighting function penalizes more strongly the higher-frequency features, the min-norm solution gives them smaller amplitudes, making the fit smoother.}
    \label{fig:fwols}
    \end{mdframed}
\end{figure}

In the Fourier case this re-weighting can be very straightforward: Each of the $p$ embedding functions $g_j(t)$ has an associated frequency $\omega_j$; we can control the fit by weighting the $j$ features by a function $f(\omega)$.
For demonstration purposes, we can choose
\begin{equation}\label{eq:f}
    f(\omega) = \frac{1}{s^2\,\omega^2 + 1}
    ~,
\end{equation}
where $s$ is a hyper-parameter controlling the (inverse) width of the weighting function in frequency space, and the frequency input will be $\omega_j$ for each feature $j$. That is,
\begin{equation}\label{eq:lambdainv}
    [\Lambda^{-1}]_{jj} = [f(\omega_j)]^2
    ~.
\end{equation}
We have chosen this form \eqref{eq:f} for $f(\omega)$ for specific reasons that will become obvious below.
The introduction of the feature weights dramatically changes the predictions; this is shown in \figurename~\ref{fig:fwols}.
High frequencies are suppressed and the prediction becomes smooth.

This all illustrates that, while OLS is affine invariant in the under-parameterized setting, the affine non-invariance in the over-parameterized setting is a property that can be exploited.
When the different features are weighted or normalized differently, the amplitudes of the components $\beta_j$ of the parameter vector $\beta$ have to change in response, which in turn changes its norm $\|\beta\|_2^2$.
That is, the details of the min-norm data-fitting parameter vector depends on the details of how the features are normalized or weighted.
In feature-weighted OLS, this property can be exploited to make the fits smooth, or meet other desiderata.

If we think of the OLS choice $\hat{\beta}$---the min-norm vector among all vectors $\beta$ that thread the data (satisfy $X\,\beta = Y$)---as the result of a kind of light regularization, then the feature weighting is an adjustment of the form of that regularization:
It asks the optimization to pull some components of $\beta$ towards zero harder than others.
In our view, feature weighting should be considered part of the investigator's choice of regularization.
The feature weighting can be thought of as altering the details of that choice, or it can be thought of as making the standard min-norm choice but in a carefully chosen, rescaled basis.

\begin{figure}[t]
    \begin{mdframed}
    \includegraphics[width=\figurewidth]{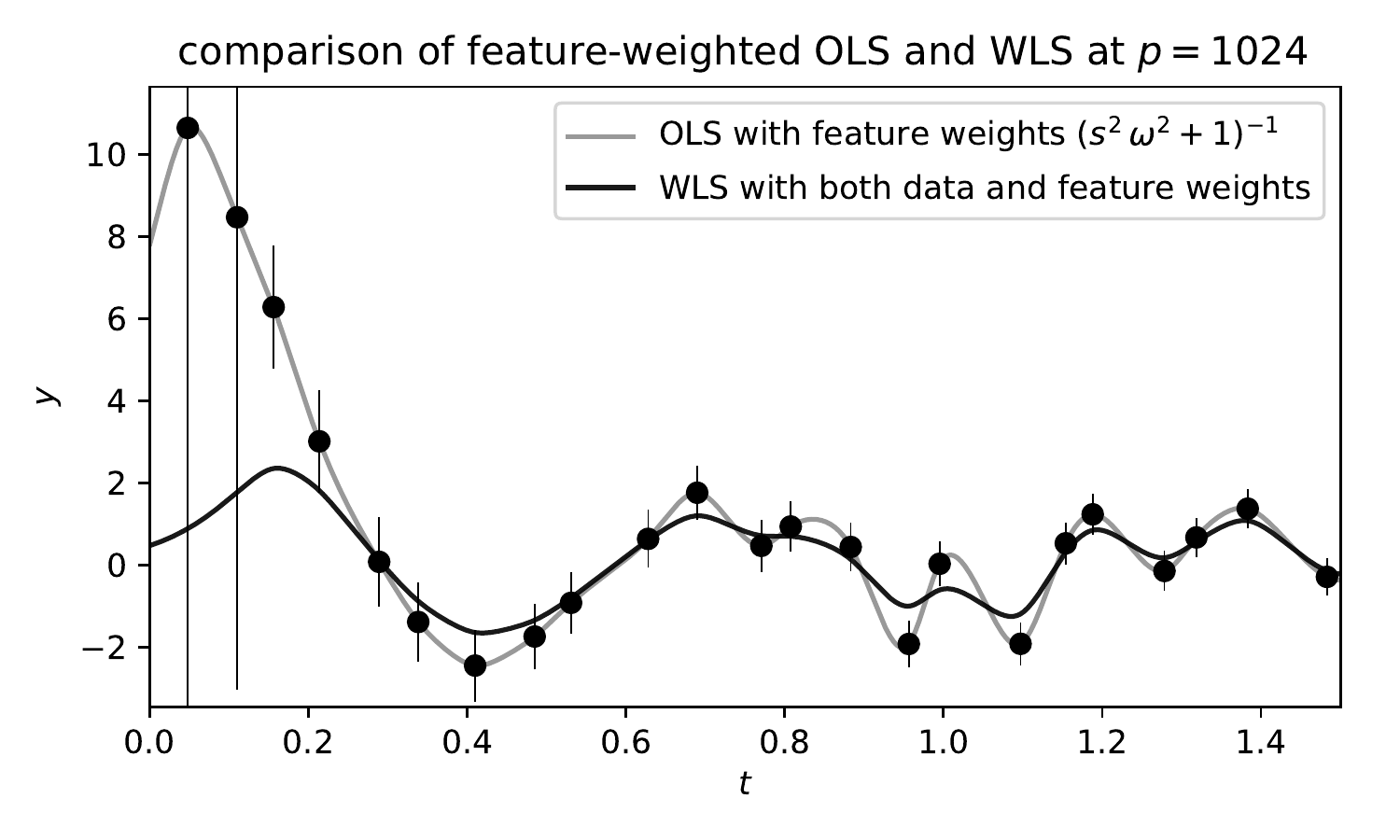}
    \caption{Comparison of the feature-weighted OLS shown in \figurename~\ref{fig:fwols} to the same but also including data weights $C^{-1}$, as in equations \eqref{eq:LambdaC1} and \eqref{eq:LambdaC2}.
    Similarly to \figurename~\ref{fig:wls}, the data weights on the diagonal of $C^{-1}$ are the inverses of the squares of the uncertainties shown as vertical error bars. The feature weights encourage the prediction to be smooth; the data weights permit it to be even smoother because they permit the prediction to miss the training data.
    Prediction results now depend on the specific amplitude or prefactor multiplying the feature weights; a particular value of $0.07\,f^2(\omega)$ was chosen for the diagonal of the feature weight matrix $\Lambda^{-1}$ for this demonstration.}
    \label{fig:fwwls}
    \end{mdframed}
\end{figure}

In the most general case, you have not just a set of feature weights $\Lambda^{-1}$, you also have a set of data-point weights $C^{-1}$ (which, in standard settings, would be the inverses of the variances of the noise affecting the data points $y_i$).
When you put these all together, the feature-weighted, data-weighted least squares predictions are given by either of these two equivalent expressions:
\begin{align}\label{eq:LambdaC1}
    \hat{Y} &= X_\ast\,(X^\top C^{-1}\,X + \Lambda)^{-1}\,X^\top C^{-1}\,Y
    \\ \label{eq:LambdaC2}
    \hat{Y} &= X_\ast\,\Lambda^{-1}\,X^\top (X\,\Lambda^{-1}\,X^\top + C)^{-1}\,Y
    ~.
\end{align}
The equivalence of these is due to the Woodbury matrix identity (\citealt{henderson1981deriving}). These expressions appear in Bayesian-inference contexts (see, for example, \citealt{products}) when $\Lambda^{-1}$ is the variance of the prior on the parameter vector $\beta$ and $C$ is the variance of the noise on the data vector $Y$.
They are solutions to the optimization
\begin{equation}
    \hat{\beta} = \arg\min_\beta \|C^{-1/2}\,(Y - X\,\beta)\|_2^2 + \|\Lambda^{1/2}\,\beta\|_2^2
    ~.
\end{equation}
The units of $C$ must be the units of $Y$ squared, and the units of $\Lambda^{-1}$ must be the square of the ratio of the units of $Y$ to the units of $X$ (the square of the units of $\beta$). These weighted-feature, weighted-data least-square forms \eqref{eq:LambdaC1} and \eqref{eq:LambdaC2} are good because (if the weightings are chosen appropriately) they lead to smooth solutions that are not required to pass precisely through every data point. This is appropriate in common, real situations in which data are noisy and the world is smooth. Which form you choose, between \eqref{eq:LambdaC1} and \eqref{eq:LambdaC2}, depends on a few things, but primarily $p/n$. If $p<n$ it's both faster and more stable to use \eqref{eq:LambdaC1}; if $p>n$ it's faster and more stable to use \eqref{eq:LambdaC2}. We show a toy example of a fit with both feature weights $\Lambda^{-1}$ and data weights $C^{-1}$ in \figurename~\ref{fig:fwwls}.

\section{How to set the number of parameters (and other hyper-parameters)}\label{sec:dd}

There is a lot of literature analyzing the performance of linear regressions as a function of the sizes $n$ and $p$, regularization strengths and forms, and so on (for example, \citealt{bartlett2020benign}, \citealt{hastie2019surprises}).
They often refer to the ``risk'', which is a statistics term for the expected squared error (mistake) made when predicting new data not in your training set.
In order to deliver values or bounds on the risk, this literature depends on knowing how the data were generated, or the family of distributions from which the data ($X$ and $Y$ in our nomenclature) were drawn.
In the Real World~(tm), you don't get this luxury.
The data are given to you without documentation!
Indeed, understanding the generating process of the world is the goal of investigations in the natural sciences (such as astronomy); the investigator does not know the generating process at the outset.

\begin{figure}[t]
    \begin{mdframed}
    \includegraphics[width=\figurewidth]{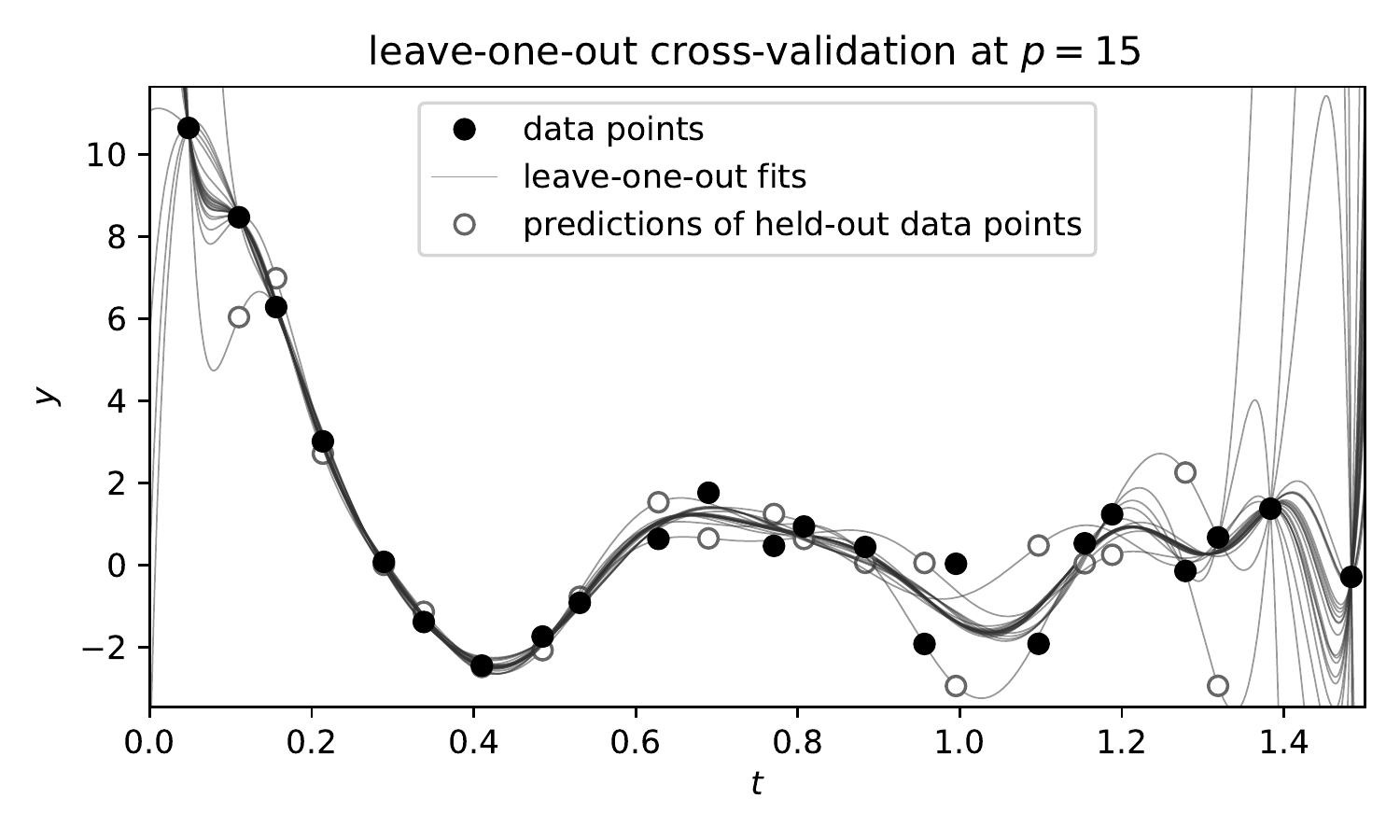}
    \caption{A demonstration of leave-one-out cross-validation for one particular model. The 23 lines are the 23 fits, for each of which one data point was held out. Also plotted on each of the 23 lines is the prediction made for that fit's held-out data point. Some of the predictions don't appear within the plot window, because the predictions obtain large amplitudes at the edges.}
    \label{fig:loo}
    \end{mdframed}
\end{figure}

Given a data set (location--data pairs $t_i, y_i$), what is the best way to empirically estimate, from those data, the out-of-sample prediction error?
That is, how do you estimate how well you are likely to predict new data?
A reasonable answer to this is cross-validation:
In cross-validation, we leave out a part of the data (in the most extreme form, leave out one single data point at a time), train the model using all but the left-out part, and predict the left-out part. Then this process is iterated over all choices of what part (or point) to leave out.
This process is illustrated in \figurename~\ref{fig:loo}, where we show $23$ fits, each of which is trained to the 22 points remaining when we leave one of the $n=23$ points out.
Also shown is the prediction, in each leave-one-out fit, for the left-out point.

\begin{figure}[t]
    \begin{mdframed}
    \includegraphics[width=\figurewidth]{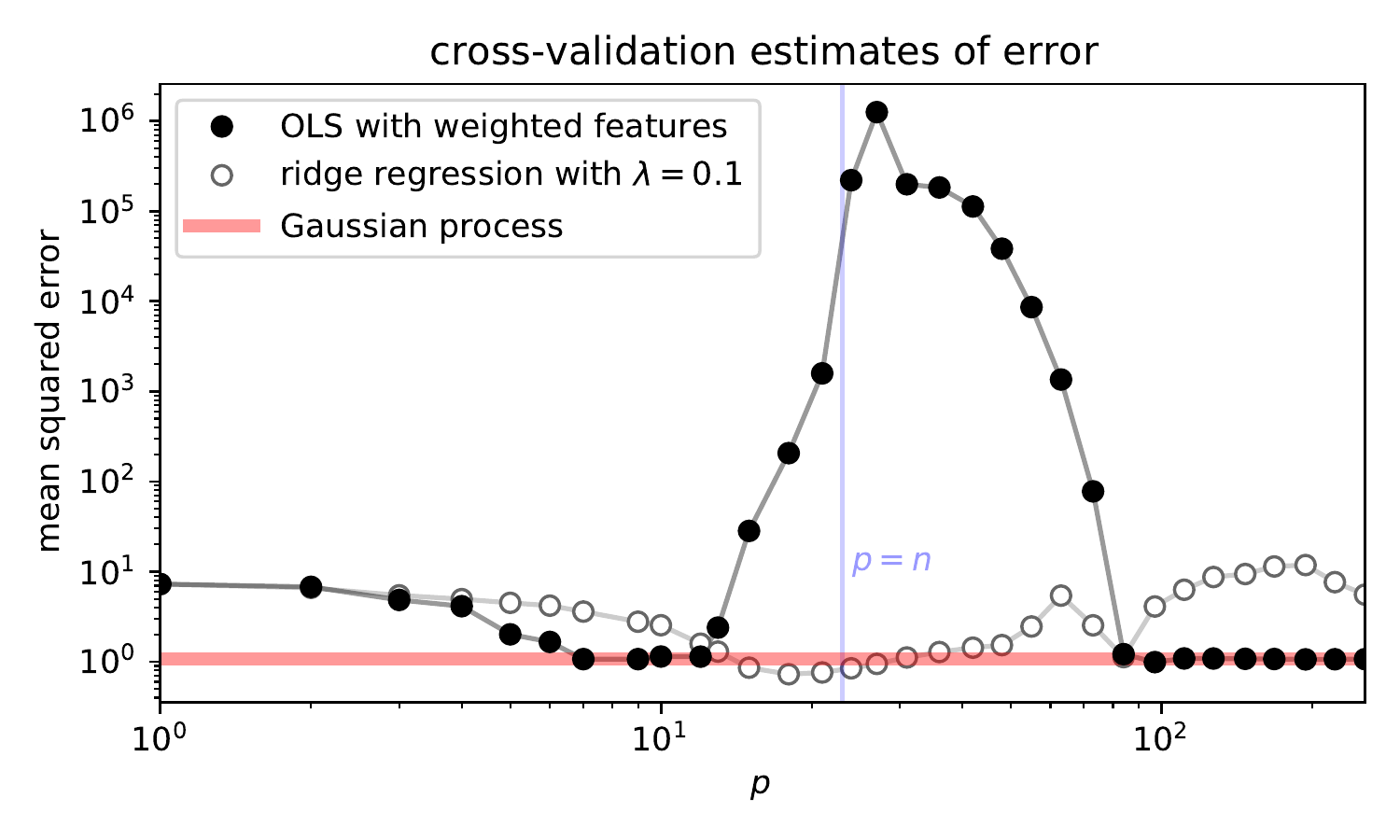}
    \caption{Leave-one-out cross-validation estimates of mean squared prediction error for the feature-weighted OLS fits, as a function of the number of features $p$. Also shown are the same for ridge regression with $\lambda=0.1$ and also the Gaussian process that we will introduce in \sectionname~\ref{sec:gp} and show in \figurename~\ref{fig:gp}; the GP result is shown as a flat line because it doesn't have an associated number $p$. In detail the first (lowest-$t$) and last (highest-$t$) data points were not used in computing the mean squared error; that choice is debatable, but we are imagining an assessment of the quality of the \emph{interpolations}, not extrapolations, of these models. Note that the OLS predictions are much, much worse at $p\approx n$ (check out all the orders of magnitude on the vertical axis) than they are at very low or very high $p$, but that ridge regression doesn't show this behavior.}
    \label{fig:cv}
    \end{mdframed}
\end{figure}

The prediction for the mean-squared error (MSE) for new data is the mean of the square of the differences between the leave-one-out predictions and the left-out data.
This leave-one-out cross-validation MSE (CVMSE) will be different for different choices for the basis, size ($p$), and regularization (including feature-weight functions) of the fits you do.
It is a fairly reliable and well-studied method for assessing predictive accuracy (see, for example, \citealt{cv}), although it does depend on some assumptions (for example, that data points are not duplicated or strongly corrrelated, and that the predictive information in the data is distributed among multiple individual data points).
An example of CVMSE for two models (the feature-weighted OLS and the ridge regression) are shown for our toy data, as a function of the number of features $p$, in \figurename~\ref{fig:cv}.
The predictive accuracy is indeed a very strong function of $p$.

In particular, the CVMSE for the feature-weighted OLS fits is bad when $p\approx n$, and much better at $p\ll n$ and $p \gg n$.
This phenomenology is not particular to this problem.
It is extremely general.
There is an effect known in many kinds of regression called ``double descent'' or ``peaking phenomenon'' or ``jamming'' in which predictive accuracy becomes very poor when the number of free parameters $p$ comes close to the number of data points $n$ (\citealt{jain198239}, \citealt{spigler2018jamming}, \citealt{geiger2019jamming}). For linear models, the ``risk''---the out-of-sample prediction error---blows up when the model capacity just becomes excessive at $p=n$ (\citealt{hastie2019surprises}).
The fundamental reason for this phenomenon is that the ordinary least-squares estimates \eqref{eq:OLS} and \eqref{eq:OLS2} require computation of the inverses $(X^\top X)^{-1}$ and $(X\,X^\top)^{-1}$ respectively, which are very badly conditioned around $p\approx n$.
This translates to a large variance for the estimator $\hat \beta$ which implies a large risk.
One way to think about this is that when the condition number of the matrix $X^\top X$ or $X\,X^\top$ is large, some directions in the data space---some linear combinations of the elements of the data vector $Y$---are very strongly amplified.\footnote{The condition number comes back up below in \sectionname~\ref{sec:implementation}. It turns out that when the condition number is large, not only do the output predictions become very sensitive to the input training data, but also the numerical (computational) stability of the linear algebra can also be badly affected. But we emphasize here that the risk goes bad when the condition number is large \emph{even if} it is possible to perform the linear algebra correctly at high precision.}
This makes the regression unreasonably sensitive to noise in the data.
The high-risk behavior at $p\approx n$ disappears with certain kinds of regularization.
These include the ridge regularization shown in \figurename~\ref{fig:cv} (also known as Tikhonov regression), but also early stopping (\citealt{hastie2019surprises}), and dimensionality reduction (\citealt{huang2020dimensionality}).

Sometimes the word ``over-fitting'' is used to describe models that are too flexible.
In our view, a model is properly called ``over-fit'' when the prediction of the regression on held-out data---what the statisticians call ``risk''---is bad.
So over-fitting happens not universally when the number of parameters is large, but instead when the number of parameters is comparable to the number of data points, and (also) the regularization is inappropriate.
\figurename~\ref{fig:cv} shows that models with both small and large numbers $p$ of parameters can make good predictions for held-out data, and that regularization can also protect a model from over-fitting, even when $p\approx n$.
We don't mean to imply that over-fitting is not a problem in regression; it can be!
Our recommendation is just that the problem of over-fitting be analyzed empirically through cross-validation, not through intuitive ideas about the number of parameters.

The CVMSE gets better and worse at different values of $p$.
That's not surprising; there are places where the fit basis and size and feature weighting are all more appropriate for your specific problem.
These good-CVMSE places are the best places to work, if you can afford to search for them and find them.
In general, if you care about predictive accuracy, it is worth doing a search for the values of $p$ (and other hyper-parameters, like regularization strengths and feature-weighting function parameters) that minimize the CVMSE.
The minimum-CVMSE choices for $p$ and the hyper-parameters will generally be very close to the choices that lead to the best predictions for new data.

That said, there is one more point to make about over-fitting, which is that, when your data set is small, it is possible to over-fit even your cross-validation.
This problem is beyond the scope of this \documentname; all we will say here is that it is not possible to make precise settings of many hyper-parameters through cross validation.
Because in cross validation you are making use of your data to set the properties of your regression, there are dangers of over-adapting your method to your data.
It is safer to have a fully independent validation data set, but this is rarely practical (and it doesn't completely protect you either; see \citealt{cifar10}).

By the way, you might have felt uncomfortable with dropping individual points in this problem, since the points are sort-of regularly spaced in the location space (the $t$ space) to begin with, and become more irregularly spaced when you leave one out.
How much does this matter?
It probably \emph{does} matter in detail, but you don't have much choice in problems like this.
All empirical measures of predictive accuracy have these kinds of problems.

\section{The Gaussian process: The limit of infinite features}\label{sec:gp}

The Gaussian process (GP; see \citealt{gpml} for a complete introduction) is a non-parametric regression that takes training data $y_i$ at coordinates $t_i$, plus a kernel function, and makes predictions $\hat{y}_\ast$ for new data at new positions $t_\ast$.
The Gaussian process mean prediction looks like%
\begin{equation}\label{eq:gpmean}
    \hat{Y}_\ast = K_\ast\,K^{-1}\,Y
    ~,
\end{equation}
where $K$ is a square $n\times n$ kernel matrix between training locations and themselves
\begin{equation}\label{eq:Kmatrix}
    [K]_{i,i'} = k(t_i,t_{i'})
    ~,
\end{equation}
and $K_\ast$ is the same except it is the rectangular kernel matrix between test locations $t_\ast$ and training locations $t_i$;
the function $k(\cdot,\cdot)$ is a positive semi-definite kernel function.
Stated this simply, the GP looks like magic; our goal here is to connect this to the feature-weighted OLS from \sectionname~\ref{sec:fwols}.

First, what does it mean for a kernel function $k(\cdot,\cdot)$ to be positive semi-definite?
It means that no matter what set of $n$ locations $t_i$ you choose, the $n\times n$ kernel matrix $K$ made from the function according to \eqref{eq:Kmatrix} has only non-negative, real eigenvalues.
The kernel function must be positive semi-definite because, among other things, it is describing the variance of a process, and variances are always non-negative.
It turns out that you can guarantee that a kernel function is positive semi-definite if you can show that it is, itself, the Fourier transform of a non-negative function (Bochner's Theorem, see for example \citealt{folland2016course}).\footnote{It is \emph{not} required that the function $k(\cdot,\cdot)$ itself be non-negative; non-negative definiteness is a different condition entirely from non-negativity.}

Although the kernel matrix $K$ is, by construction, always positive semi-definite, it can have extremely bad or even infinite condition number. That is, it can lead your code or implementation into very unstable linear-algebra operations. We discuss how to handle these issues below in \sectionname~\ref{sec:implementation}.

Above, we called the GP prediction the ``mean''. This is because the Gaussian process is a model for a mean \emph{and variance} in function space.
In addition to the mean $\hat{Y}_\ast$ predicted\footnote{There is an additional point worthy of mention here, which is that the expression \eqref{eq:gpmean} is implicitly for a GP with a zero prior mean. There is a more general expression that subtracts a prior mean function from the $Y$ values and adds the prior mean back in to the $\hat{Y}_\ast$ values. See \citet{gpml} for all the math.} in \eqref{eq:gpmean}, the GP also predicts a \emph{variance} in the $Y_\ast$ space around that mean.
In this \documentname, we are going to treat the GP as producing only a mean prediction, where the shape of the kernel function matters, but the amplitude of the kernel function does not (the prediction is a ratio of kernel matrices, so the kernel amplitudes cancel).
However, in Bayesian-inference contexts the amplitude of the GP kernel is important, and the predicted variances are important.
When we consider only the mean prediction of the GP, as we do here, then the GP is a kind of linear filter that operates on data $Y$ and predicts or interpolates to new data $Y_\ast$.
Classically, this linear filter is sometimes called a Wiener filter (more on this in a moment) or \emph{kriging} (possibly because of \citealt{krige}).

Importantly for us, there is a strong connection between the feature-weighted OLS and the GP.
In particular, when we take the limit of infinite features ($p\to\infty$; provided the limit exists) we get kernel matrices $K$ in place of the $X\,\Lambda^{-1}\,X^\top$ matrix products:
\begin{align}
    \lim_{p\to\infty} X\,\Lambda^{-1}\,X^\top &= K
    \\
    \lim_{p\to\infty} X_\ast\,\Lambda^{-1}\,X^\top &= K_\ast
    ~,
    \label{eq.limit}
\end{align}
where $\Lambda^{-1}$ is the diagonal matrix of weights, and element $i,i'$ of $K$ is obtained by evaluating a kernel function $k(t_i,t_{i'})$.
Equivalently, the limit is
\begin{equation}
    \lim_{p\to\infty} \sum_{j=1}^p [\Lambda^{-1}]_{jj}\,g_j(t)\,g_{j}(t') = k(t, t')
    ~,
\end{equation}
where we have used the diagonality of $\Lambda$ to make a single sum over $j$.
The specific form of the kernel function $k(\cdot,\cdot)$ depends on the basis (the features) we choose, and the weighting of the basis functions in the OLS.

The connection between the infinite basis chosen (the form and weighting of the features) and the kernel function is governed by Mercer's theorem (see \citealt{minh2006mercer}).
However, if the basis is Fourier, as we chose above in \eqref{eq:basis}, and the spacing between modes ($\Delta\omega =\omega_{j+2}-\omega_j$) is small enough, the kernel approximates the Fourier transform of the square of the weighting function $f(\omega)$ we use to weight the features.

\begin{figure}[t]
    \begin{mdframed}
    \includegraphics[width=\figurewidth]{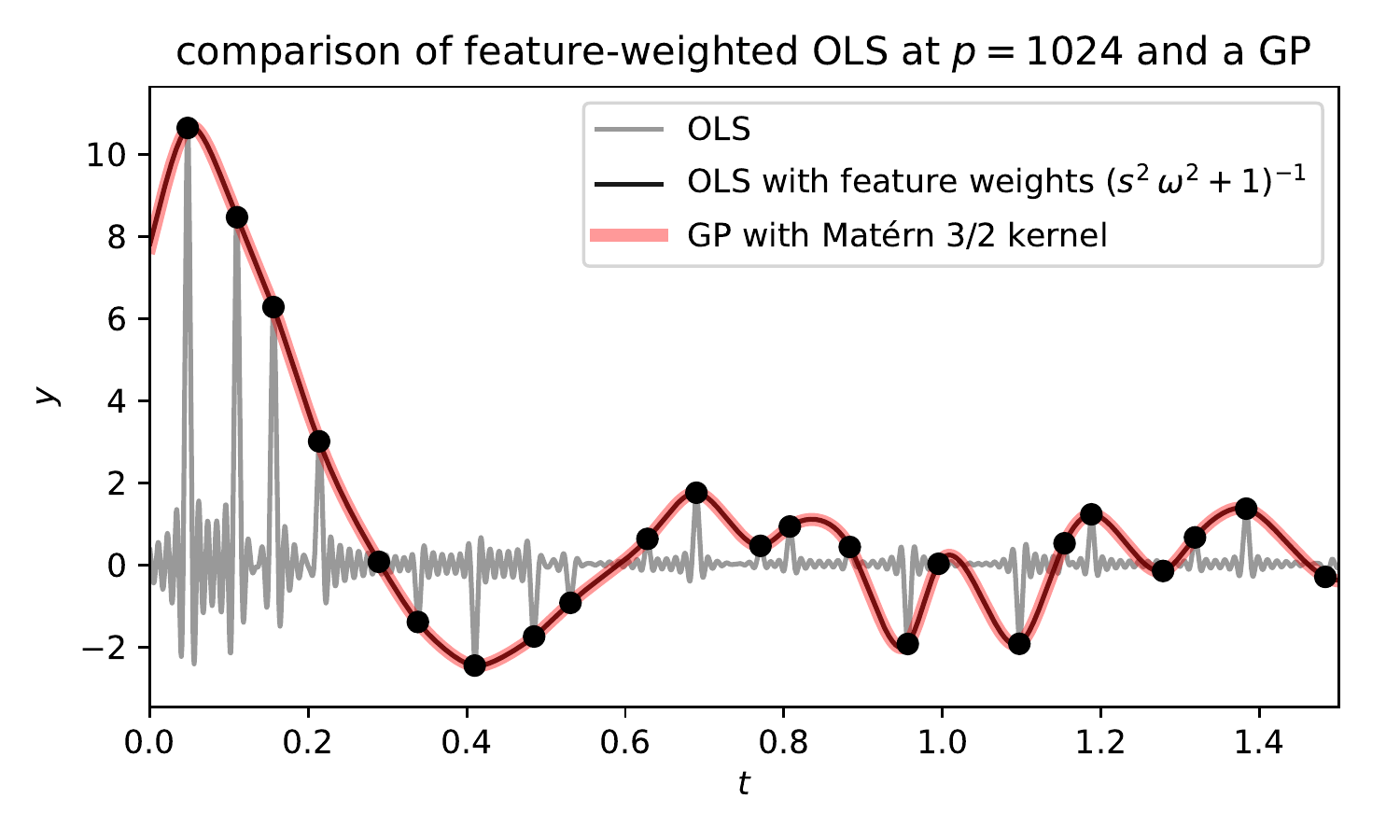}
    \caption{Comparison of feature-weighted ordinary least squares with a Gaussian process. The figure shows the OLS fit, the feature-weighted version with the particular feature weighting given in \eqref{eq:f}, and the GP fit using the kernel \eqref{eq:F} that is the Fourier transform of the square of the feature weighting. The feature-weighted fit and the GP are essentially identical, as expected.}
    \label{fig:gp}
    \end{mdframed}
\end{figure}

That is, in the case of the specific example of weight function $f(\omega)$ given in equation \eqref{eq:f}, we can connect our feature-weighted OLS to an equivalent GP if we know the Fourier transform of the square of $f(\omega)$. We chose that specific form for $f(\omega)$ because it has a square that is a member of a Fourier-transform pair:
\begin{align}
    \FT[F(t)] &= [f(\omega)]^2
    \\ \label{eq:F}
    F(t) &= \sqrt{\frac{\pi}{8}}\,\left(1 + \frac{|t|}{s}\right)\,\exp -\frac{|t|}{s}
    ~.
\end{align}
This latter function is also known as the Mat\'ern $3/2$ kernel function; it will become the kernel function for the Gaussian process when we take the limit.
In the limit $\Delta\omega\to 0$ the specific example of the $p\to\infty$ feature-weighted OLS given here becomes a GP with kernel function $F(\cdot)$ (under the mild assumptions that guarantee that the discrete Fourier transform converges to the continuous Fourier transform, see \citealt{ft}):
\begin{align}
    k(t_i,t_{i'}) &\to F(t_i-t_{i'})
    \\ \label{eq:k}
    &= \sqrt{\frac{\pi}{8}}\,\left(1 + \frac{|t_i - t_{i'}|}{s}\right)\,\exp -\frac{|t_i - t_{i'}|}{s}
    ~.
\end{align}
Technically, the kernel function $k(\cdot,\cdot)$ converges to the Fourier transform of the square of the weighting function $f^2(\cdot)$ in the limit $\Delta\omega\to 0$ (the definition of $k(\cdot,\cdot)$ already assumes $p\to \infty$).
However, provided that the spacing of the Fourier modes in frequency space is $\Delta\omega\ll 1 / \Delta t_{\max}$ and the maximum frequency is $\floor{p / 2}\,\Delta\omega \gg 1 / \min(s, \Delta t_{\min})$, where
\begin{align}
    \Delta t_{\min} &\equiv \min_{i\ne j}|t_i - t_j|
    \\
    \Delta t_{\max} &\equiv \max_{i,j}|t_i - t_j|
    ~,
\end{align}
it will be true that the OLS with feature weights $f(\omega)$ will closely approximate the mean of a GP with kernel $k(\Delta t) = \FT[f^2]$ (\citealt{ft}).
The comparison of the OLS and GP is shown in \figurename~\ref{fig:gp}.

In our toy examples, we use the feature weighting that generates the Mat\'ern 3/2 kernel.
It is important to emphasize that there are literally infinite alternative choices that can be made here, and hundreds even if you restrict to kernels with known closed forms.
Indeed, any function $F(t-t')$ that has a finite, all-positive Fourier transform $f^2(\omega)$ can be substituted for the Mat\'ern kernel and associated frequency weighting function we use here.

The particular kernel we obtain in equation \eqref{eq:k} is a stationary kernel, meaning it depends only on absolute values of time differences $|t - t'|$. Not all $p\to\infty$ kernels will be stationary. The limit $p\to\infty$ of $X\,\Lambda^{-1}\,X^\top$ in \eqref{eq.limit} leads to a stationary kernel $k(\cdot,\cdot)$ in this case because the feature embedding in $X$ is a set of sines and cosines.
Sines and cosines form a basis for translation-invariant function spaces.
If we had made a different choice for our basis, such as polynomials or wavelets, we would have obtained a non-stationary kernel in the $p\to\infty$ limit.

At the end of \sectionname~\ref{sec:fwols}, we discussed including not just feature weights in a matrix $\Lambda^{-1}$ but also data weights in a matrix $C^{-1}$. The GP also permits this, and it is often a very good idea. The generalization of equations \eqref{eq:LambdaC1} and \eqref{eq:LambdaC2} to the GP case is 
\begin{equation}\label{eq:gpC}
    \hat{Y}_\ast = K_\ast\,(K + C)^{-1}\,Y
    ~.
\end{equation}
As before, in contexts where you have independent uncertainties on each element $y_i$ of your training data $Y$, $C^{-1}$ would naturally be set to the diagonal matrix containing the inverses of the variances of those uncertainties (so $C$ would contain the variances).
This form \eqref{eq:gpC} is used a lot in astronomy and cosmology (for example, \citealt{zaroubi}, \citealt{aigrain}, \citealt{celerite}) and it is also the standard form given for the Wiener filter, where the kernel function generating $K$ is the Fourier transform of the ``signal power'' and the kernel function generating $C$ is the Fourier transform of the ``noise power''.
In this form, the amplitude of the kernel function matters, because it is competing, in some sense, the variance $K$ of the GP against the variance $C$ of the noise.
It is not sufficient to get the shape of the kernel function right to make a stable prediction of the mean, when using form \eqref{eq:gpC}.

\section{Uncertainties on the predictions}\label{sec:uncertainty}

Often you need to compute not just a prediction, but also an uncertainty on that prediction.
How faithfully you must compute that uncertainty will depend strongly on the context in which you are doing the fitting or interpolation.
However, it is often the case in the physical sciences that predictions are required to come with good or conservative estimates of uncertainty.
There are four-ish sources of uncertainty in the predictions you are making in problems like these:
(1)~The data points $y_i$ you have are individually noisy.
(2)~There are finitely many of those data points (there are $n$ of them) and there are gaps between them in the location space $t$.
(3)~The data points might have uncertain locations or location measurements $t_i$.
(4)~And the predictions you make depend on hyper-parameter choices, such as the form of the basis, the number of parameters $p$, and any regularization or feature weighting.
These four different sources of uncertainty propagate differently and are differently ``simple'' to deal with.
In particular it turns out that the easiest sources of uncertainty to understand and propagate are those coming from (1)~the noise in the $y_i$ and (2)~the number $n$ and locations $t_i$ of the data.
The uncertainties coming from (3)~uncertainties in the locations $t_i$, and (4)~model choices, are both much harder to model and propagate.

In the best-case scenario, you might have accurate estimates of the variances $[C]_{ii} = \sigma_i^2$ of the noise contributions affecting your training data values $y_i$, the noise might be Gaussian with zero mean, you might have locations $t_i$ that are extremely accurately known, and you might have justifiable prior variances $[\Lambda^{-1}]_{jj}$ on your parameters $\beta_j$.
In this case, you can assume that the linear model with $p$ parameters is a good model for your data, and write down a likelihood function and a prior.
In this case, you can turn the Bayesian crank and deliver a posterior density for the predicted values $Y_\ast$ at the test locations $t_\ast$.
The second derivative of the natural logarithm of the posterior density can then be processed into a standard error on the predictions $\hat{Y}_\ast$ (or you can deliver full posterior densities somehow).
The details of this are way beyond the scope of this \documentname, but this (or a closely related) problem is discussed in some detail by us elsewhere (\citealt{products}).

That scenario is best-case, but it does involve a lot of assumptions.
That is, if you don't trust your estimates of the noise variances, or if you think the noise might be non-Gaussian, then the likelihood approaches will give poor uncertainty estimates.
In this sense, it is more conservative to make use of more empirical methods for uncertainty estimation.
The leading candidates are jackknife or bootstrap resamplings (for example, \citealt{bootjack}).
Here we will focus on jackknife.

\begin{figure}[t]
    \begin{mdframed}
    \includegraphics[width=\figurewidth]{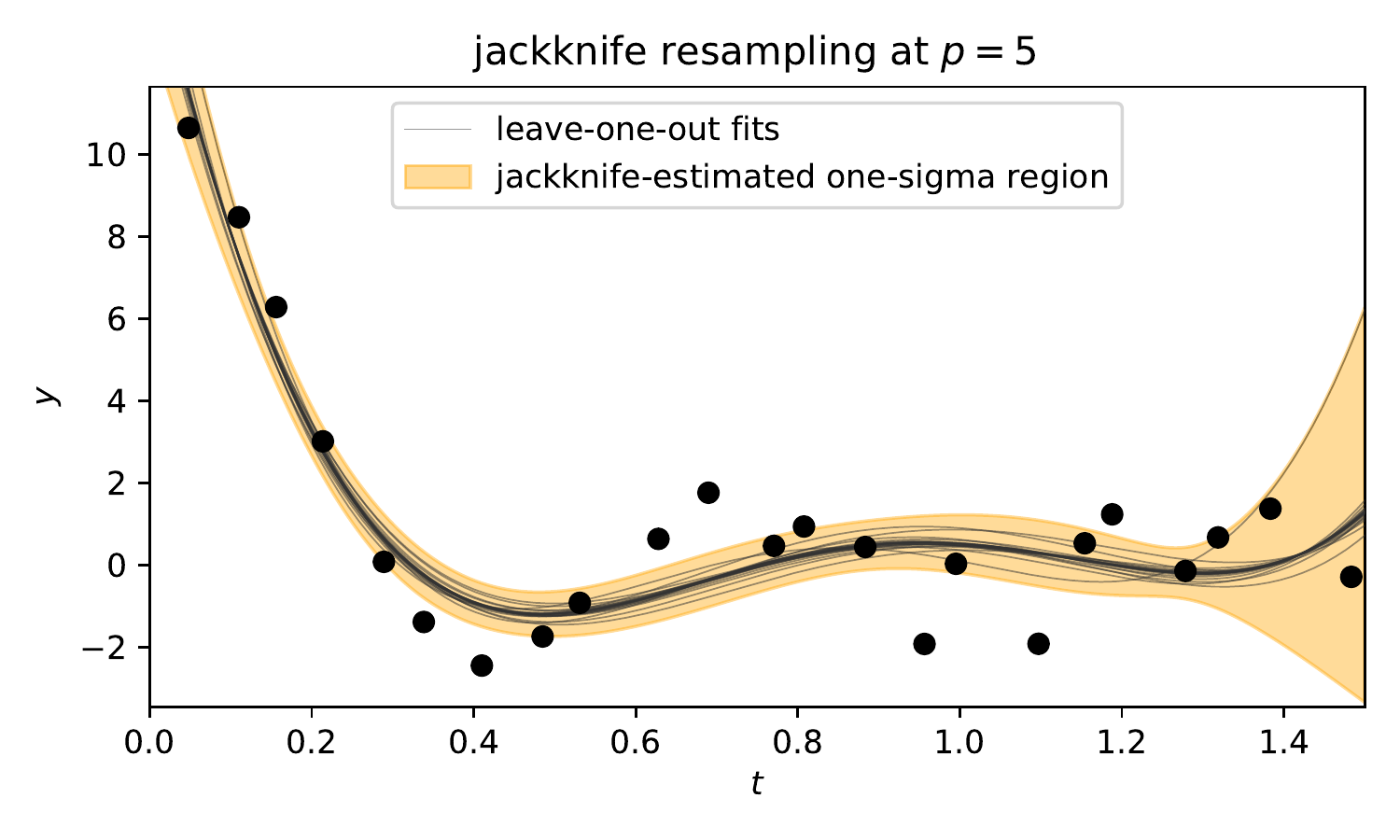}
    \caption{The jackknife-estimated uncertainty on an ordinary least squares fit. What is shown is the diagonal of the empirical jackknife estimate $\hat{\Sigma}$ of the variance tensor on the prediction $\hat{Y}_\ast$ of an OLS fit. In other words, the off-diagonal entries of $\hat{\Sigma}_\ast$ are being ignored here; there are important covariances that are hard to visualize (and beyond our scope). The one-sigma region is larger than the span of the jackknife trials; this is because the jackknife trials are strongly correlated; the jackknife formula \eqref{eq:jackknife} compensates for those correlations.}
    \label{fig:uncertainty}
    \end{mdframed}
\end{figure}

The idea of jackknife resampling is that we make $k$ subsamples of the data, in each of which we have dropped (or held out) a unique fraction $1/k$ of the data.
The variance of the results (in this case the prediction $\hat{Y}_\ast$) across the jackknife samples can be transformed into an estimate of the variance of the estimator acting on the full data set.
In the most extreme form---leave-one-out jackknife---we set $k=n$.
In this case, the jackknife estimate $\hat{\Sigma}_\ast$ of the variance of the best-fit predictions $\hat{Y}_\ast$ is given by
\begin{equation}\label{eq:jackknife}
    \hat{\Sigma}_\ast = \frac{n-1}{n}\,\sum_{i=1}^n (\hat{Y}_{\ast i} - \hat{Y}_\ast)\,(\hat{Y}_{\ast i} - \hat{Y}_\ast)^\top
    ~,
\end{equation}
where $\hat{Y}_{\ast i}$ is the estimate of $Y_\ast$ made after \emph{leaving out} the one training point $t_i, y_i$, and $\hat{Y}_\ast$ is the estimate made using \emph{all} the data.
The pre-factor $(n-1)/n$ in \eqref{eq:jackknife} is not anything like the $1/(n-1)$ that you use when you estimate a standard variance; this difference comes from the fact that the jackknife subsamples are extremely correlated; the pre-factor is computed to amplify the jackknife variance into an unbiased estimate of the variance of the estimator $\hat{Y}_\ast$.
The square-root of the trace of this uncertainty variance $\hat{\Sigma}$ (that is, what you might call the jackknife estimate of the standard error) is shown in \figurename~\ref{fig:uncertainty}.
Note the similarity between jackknife and cross-validation; it is relevant and important: These empirical estimates of uncertainty and prediction error are closely related.

These methods---full likelihood analysis and jackknife---take into account (1)~the noise in the data $y_i$ and (2)~the number and spacing of the training data points.
They do not, by themselves, take into account (3)~the uncertainties on the locations $t_i$.
There are no simple methods for uncertainty propagation from the locations $t_i$ into the predictions $\hat{Y}_\ast$.
One option is to resample the $t_i$ according to your uncertainty estimates, and re-do the fits.
That's potentially expensive, and overly conservative (because it ignores the information about the $t_i$ coming from the $y_i$).
Another is to linearize the first derivative of the best-fit curve $y(t)$ and propagate uncertainty using those derivatives.
This is not conservative, because it only works if the location uncertainties are small relative to substantial changes in the slope of the predictions with $t$.
The most extreme option would be to \emph{simultaneously fit} for the prediction $\hat{Y}_\ast$ and \emph{all} of the noisily measured $t_i$.
That's a great idea, but that fit would be extremely computationally expensive, and no reasonable fit objective would be convex.
All of these ideas are out of scope here.

Finally, in order to take account of (4)~the uncertainty coming from your choices of basis, number $p$, and regularization, you might have to know (or learn) the distribution over \emph{these} things.
Since these choices are hyper-parameters, any inference that propagates uncertainties coming from these choices would have to be hierarchical in structure.
That is way, way out of scope.

\section{Implementation notes}\label{sec:implementation}

All of the code used to make the figures for this \documentname\ is available publicly.\footnote{\url{https://github.com/davidwhogg/FlexibleLinearModels}}
Although the examples are toys, the implementation of everything can be generalized for real-data situations.
In our code, there are some aspects of the linear-algebra implementation that might seem odd.
Here are some comments.

Mostly, linear algebra stability comes down to the \emph{condition number} of the matrix in question.
The condition number, for our purposes, is the ratio of the largest eigenvalue (for non-negative definite matrices) to the smallest \emph{non-zero} eigenvalue.
For rectangular feature matrices it is the ratio of the largest singular value (that is, what you get out of a singular value decomposition) to the smallest nonzero singular value.

When the condition number is large, different linear-algebra approaches will differ.
For example, the function call \code{solve(A, b)} should give you the best estimate it can for the product $A^{-1}\,b$, whereas the function call \code{dot(inv(A), b)}, which is the same on paper, will give you the dot product between \code{b} and the best estimate that \code{inv()} can find for the inverse of \code{A}.
When the condition number of $A$ is large, these two values can be very different, and the \code{solve()} will be better.\footnote{This claim contradicts what is stated in the abstract of \citet{solve}, but this claim is based on our real numerical experiments on real matrices, so we stand by it. What is uncontroversial is that \code{solve(A, b)} will always perform non-worse than \code{dot(inv(A), b)}; sometimes it will perform far better.}
Use \code{solve(A, b)} and never \code{dot(inv(A), b)} unless there are compelling code-structure reasons to use the latter, such as repeated calls (but even then, a Cholesky decomposition followed by repeated Cholesky solves is probably a more stable solution).

If your condition number gets very large, even \code{solve()} can give bad results, because very small eigenvalues of the matrix are being poorly estimated and then inverted.
That's unstable.
It is better to zero out those small eigenvalues---it is better to destroy bad information than to use it---and perform a pseudo-inverse in those cases.
So if you really want to be safe (and you do, here) you should really use \code{lstsq(A, b, rcond=tiny)} instead of \code{solve(A, b)}.
The \code{lstsq()} function requires a \code{rcond} input, which says at what (dimensionless) precision to zero-out low eigenvalues.
It usually makes sense to set this dimensionless \code{rcond} input to something like machine precision, which is about \code{1e-16} for our current hardware--software setups.

You might think that all of this is academic, but it really isn't when the number of features $p$ is close to the number of data points $n$.
For example, in our toy-data OLS experiments in this paper with $n=23$, the matrix $X^\top X$ has a condition number (ratio of largest eigenvalue to smallest nonzero eigenvalue) that saturates machine precision for the entire range $12 < p < 50$.
And that's for $n=23$; things generally get worse as $n$ gets larger.

Related to this, if you have a choice between formulations that involve an inverse of a square, like $(X^\top X)^{-1}\,X^\top Y$, and formulations that involve a pseudo-inverse, like $X^\dagger\,Y$, you should do the latter, because $X^\top X$ has the \emph{square} of the condition number of $X$.
When you want to execute $X^\dagger\,Y$, as we do in \eqref{eq:OLS3}, you should again use \code{lstsq()}, which was literally designed for these applications; use \code{lstsq(X, Y, rcond=tiny)}.
This function returns the best estimate it can for $X^\dagger\,Y$, so
it should be better than computing a pseudo-inverse \code{pinv(X)} and then matrix-multiplying \code{pinv(X)} with \code{Y}.
Again here it usually makes sense to set this \code{rcond} input to something like machine precision.

Once $p$ is large enough, if you are using a feature weighting with $\Lambda^{-1}$ with diagonal entries $[\Lambda^{-1}]_{jj}$ that decrease to zero with $j$, at some point, at machine precision, the additional columns you are adding to $X$ are effectively all zeros.
They will literally underflow the linear-algebra representation.
That's not a problem if you implement your linear algebra well (that is, use \code{lstsq()} appropriately), but it does mean that your predictions and cross-validations will saturate at some $p$ (as we see them do in \figurename~\ref{fig:cv}).

In multiple places, but especially \eqref{eq:LambdaC1} and \eqref{eq:LambdaC2}, you are performing operations on matrices that are diagonal ($C$ and $\Lambda$ and their inverses are all diagonal).
You should avoid ever constructing diagonal matrices.
You can multiply $X$ or $X^\top$ by a diagonal matrix by just multiplying the rows (or columns) by the diagonal entries.
And you can invert by just inverting the diagonal entries.
Avoid constructing and operating on operators that are almost entirely zeros, unless you have a very efficient sparse linear algebra implementation and you are a power user.

And finally, it is most numerically stable to operate on the smallest matrices you can.
For example, when you have the choice between the formulation in equation \eqref{eq:LambdaC1} and that in equation \eqref{eq:LambdaC2}, you should choose the former when $p<n$ and the latter when $p>n$.
That way you are always doing the heavy linear algebra (\code{solve()} and \code{lstsq()} function calls) at a size $\min(p, n)$, which is both faster and more stable than linear algebra at $\max(p, n)$. And it is much faster when you make these choices correctly:
Linear algebra scales naively as the dimension \emph{cubed}.
(In practice---with excellent packages---it actually scales with a power more like 2.6 than 3, but still!)

\section{Discussion}\label{sec:discussion}

This \documentname\ was about linear fitting with very flexible models, for interpolation, prediction, and de-noising of data.
We encouraged you to consider using very big models, but being intentional about regularization.
These settings (over-parameterized, but carefully regularized) are adaptive and useful, and they connect, as we showed, to Gaussian processes in the limit, which are well-studied workhorses of machine learning and data science.

All the rage these days, is instead \emph{nonlinear fitting}, with \emph{deep learning} and the like (\citealt{deep}, for example).
We didn't discuss any of that.
However, many of the high-level lessons in this \documentname\ carry over:
All of these tools work well when the function space is flexible but also carefully regularized.
In these nonlinear settings, regularization takes many additional new forms, like early stopping (\citealt{earlystop}), dropout (\citealt{dropout}), and restricted network structure (such as convolutional; see for example \citealt{bishop}).

We only considered the setting in which you care about predicting new data $Y_\ast$ and never the setting in which you care about the internal parameters $\beta$ of the linear fit.
However, if you are, say, measuring the power spectrum of a process (as we are, often, in cosmology), then you care about these amplitudes themselves.
In this case, any regularization you apply takes the role of a prior or prior information.
That prior must be chosen (in those cases) with great care, because the answers you get will be influenced by that choice; when it doesn't properly conform to your true beliefs, your answers will be distorted in bad ways.
And, technically, when you take this view (that the parameters $\beta$ matter), you also have to get serious about the noise model on the data.
That is, the beliefs encoded in the data noise variance tensor $C$ also must represent your true beliefs if you don't want your answers distorted in bad ways.
All of that is out of scope here, but we say things about it all elsewhere (for example, \citealt{fitting}).

In many cases, when the goal is just to interpolate or de-noise data, investigators use running means (or medians), low-pass filters, or explicit interpolators (like cubic spline interpolation).
These methods all have close relationships with what is written here.
Indeed, a running mean, a low-pass filter, and a cubic-spline interpolator can all be written as a linear operator (the details of which depends on the locations $t_i$ of the data) acting on the input data vector $Y$, just like our linear fits produce linear operators.
This means that in many cases, these methods can be translated into versions of the methods we have presented in this \documentname.
A full translation is beyond our scope, though.
And our view is that the value of making explicit investigator choices about basis and the regularization makes the linear fitting approach better in general.

Regularization was perhaps the biggest theme of this \documentname.
But we only really considered variants of L2 regularization (or ridge or Tikhonov).
There are other regularizations, even other convex regularizations.
A valuable and useful option is L1 or the lasso (\citealt{lasso}), which encourages sparsity---it encourages parameters $\beta_j$ to take the value $\beta_j=0$ exactly, where possible.
This kind of regularization makes sense when we have prior beliefs that the functional forms we seek will in fact be sparse in our chosen basis.
Usually this kind of consideration is less important in interpolation, prediction, and de-noising settings, but it is not unheard of.
A nice recent result is that the feature weighting that we employ here can be used with L1 regularization to obtain simultaneously smoothness \emph{and} sparsity (\citealt{rauhut2016interpolation}).

We discussed deterministic, ordered expansions like Fourier series and polynomials, but there is another class of \emph{random features} methods (see, for example, \citealt{rahimi2007random}) that we did not discuss.
Briefly, the idea with random features is that instead of weighting features that are regularly spaced in frequency space with a weighting function $f(\omega)$, we could have generated unweighted features but randomly from a probability distribution $\propto f^2(\omega)$.
The same Gaussian process limit appears as $p\to\infty$ (provided we choose those features with appropriately random phases too).
These random-feature approaches have rarely been used in the natural sciences, but they are potentially of interest in many applications.

In the toy examples used throughout this \documentname, we considered data with locations $t_i$ in a one-dimensional location space or ambient space.
This was a choice for simplicity of visualization; in principle the locations $t_i$ could be higher dimensional, or live in a different kind of space.
The most important consideration about the dimensionality and range of the locations is that the functions $g_j(t)$ have to sensibly take the locations $t_i$ as input.
In astronomy and cosmology contexts, it is common for the locations $t_i$ to be positions on the sphere, and it is common for the natural basis functions to be spherical harmonics, for example.

The methods in this \documentname\ are all \emph{discriminative}, in the sense that we took the locations $t_i$ to be prior or primary;
the goal was to find a function of locations $t_i$ that predicts data $y_i$.
This asymmetry between the locations and the data led to concerning statements in \sectionname~\ref{sec:uncertainty} when we considered the possibility that the locations $t_i$ themselves might be noisy or uncertain.
An alternative formulation to these discriminative models are \emph{generative} models, in which the model attempts not just to predict the $y_i$ from the $t_i$ but instead predicts both the $t_i$ and the $y_i$.
A model that generates both can be used to make predictions $y_\ast$ at new locations $t_\ast$ by executing an inference or inverse problem on the generative model.
Generative models are generally non-convex---and harder to execute---if you want the relationship between the $t_i$ and the $y_i$ to be nonlinear (for an example of this in astronomy, see, for example, \citealt{thecannon}).
In these cases you don't get closed-form solutions, and there aren't known guarantees (like the Gauss--Markov theorem) about performance.
But they are more general, and often more appropriate, especially when both the locations $t_i$ and the data $y_i$ are comparably noisy measurements.

In our toy examples, we used a Fourier series.
That is just one choice among many.
However, \emph{it is often a great choice}.
When you choose the Fourier basis (and---importantly---for every cosine term you include the corresponding sine term), the matrix $X\,\Lambda^{-1}\,X^\top$ (and the limiting kernel matrix $K$ at $p\to\infty$) has the property that every matrix element $[X\,\Lambda^{-1}\,X^\top]_{ii'}$ depends on (or can be calculated from) just the absolute difference $|t_i-t_{i'}|$.
That is, in this basis, all fitting methods become technically \emph{stationary}.
This all relates to the translation-independence properties of the Fourier basis.
So although the Fourier basis is just one choice among many, it is the \emph{right choice} when you think your problem has (or might have) certain kinds of translation invariances.


{\raggedright
\setlength{\bibsep}{0pt plus 0.3ex} 
\bibliographystyle{authordate1}
\bibliography{references}
}

\appendix
\section{Optimization arguments}\label{app:math}

As mentioned in the main text, the optimal solutions of the unconstrained objectives \eqref{eq:opt1} and \eqref{eq:ridge} can be obtained by computing the first-order critical points. The first-order critical points are global minima because all the objectives considered are convex in the regression coefficients $\beta$.

To derive the optimal solutions of the constrained optimization \eqref{eq:opt2} and similar, we use a classical result in quadratic optimization (see for instance \citealt{nocedal2006numerical}, Ch.~16):
Consider the quadratic optimization problem 
\begin{equation} \label{qp}
    \arg\min_x \frac{1}{2} x^T G\, x + x^\top c \quad \text{ subject to } A\,x=b
    ~.
\end{equation}
The first-order necessary conditions for $x^*$ to be a solution of \eqref{qp} is that there is a vector $\lambda^*$ (known as Lagrange multipliers) such that the following system of equations is satisfied:
\begin{equation}
    \left[
    \begin{matrix}
    G & A^\top \\ 
    A & 0
    \end{matrix}\right]
    \left[
    \begin{matrix}
    x^* \\ 
    \lambda^*
    \end{matrix}\right] = 
    \left[
    \begin{matrix}
    -c \\ 
    b
    \end{matrix}\right]
    ~.
    \label{kkt}
\end{equation}
Equation \eqref{kkt} is typically known as the Karush-Kuhn-Tucker conditions (KKT). If the objective function in \eqref{qp} is convex, the KKT conditions are also sufficient for optimality.\footnote{The KKT conditions can also characterize optimality in more general cases. For instance, if $A$ has full row rank and $Z$ is the nullspace of $A$, if $Z^\top G\, Z$ is positive semidefinite the KKT conditions are sufficient for optimality. If $Z^\top G\, Z$ is positive definite we also have that the solution of \eqref{qp} is unique (which is typically the case in the underparameterized linear regression but not in the overparameterized).}

Using this formulation it is easy to check that the solution of \eqref{eq:opt2} leads to predictions in \eqref{eq:ridge_sol}, and the solution of \eqref{eq:weighted} produce \eqref{eq:weighted_sol}. For instance, in order to show the latter we consider 
\begin{equation} \label{eq:weighted_app}
    \hat{\beta} = \arg\min_\beta \|\Lambda^{1/2}\,\beta\|_2^2 ~~\mbox{subject to}~~ Y = X\,\beta
    ~,
\end{equation}
or equivalently 
\begin{equation}
    \hat{\beta} = \arg\min_\beta \frac{1}{2}\,\beta^\top \Lambda\,\beta ~~\mbox{subject to}~~  X\,\beta = Y
    ~,
\end{equation}
obtaining the formulation in \eqref{qp} for $G=\Lambda$, $c=0$, $A=X$, $b=Y$ and $x=\beta$. Using the KKT conditions we obtain
\begin{equation}
    \left[
    \begin{matrix}
    \hat \beta \\ 
    \lambda^*
    \end{matrix}\right] =
    \left[
    \begin{matrix}
    \Lambda & X^\top \\ 
    X & 0
    \end{matrix}\right]^{-1}
    \left[
    \begin{matrix}
    0 \\ 
    Y
    \end{matrix}\right]
    ~,
    \label{kkt2}
\end{equation}
(assuming the KKT matrix is invertible). Luckily there exists a complete characterization for the inverse of $2\times 2$ block matrices (see, for example, \citealt{lu2002inverses}). In particular if $A$ and $D - CA^{-1}B$ are invertible matrices we have
\begin{multline} 
 {\begin{bmatrix}  {A} &   {B} \\  {C} &  {D} \end{bmatrix}}^{-1}= \\ {\begin{bmatrix}  {A} ^{-1}+  {A} ^{-1}  {B} \left(  {D} -  {CA} ^{-1}  {B} \right)^{-1}  {CA} ^{-1}& -  {A} ^{-1}  {B} \left(  {D} -  {CA} ^{-1}  {B} \right)^{-1}\\-\left(  {D} -  {CA} ^{-1}  {B} \right)^{-1}  {CA} ^{-1}& \left(  {D} -  {CA} ^{-1}  {B} \right)^{-1}\end{bmatrix}}
 ~.
\end{multline}

In the particular case of \eqref{kkt2} it suffices to compute the top right block the inverse. If both $\Lambda$ and $X\,\Lambda^{-1}X^\top$ are invertible we obtain 
\begin{equation}
    \hat \beta = \Lambda^{-1}\,X^\top(X\,\Lambda^{-1}X^\top)^{-1}\,Y
    ~. 
\end{equation}

\end{document}